\documentclass{jfm}

\usepackage{graphicx}
\usepackage{epstopdf, epsfig}
\usepackage{amsmath}
\usepackage{mathtools}
\usepackage{subfigure}
\usepackage[export]{adjustbox}
\usepackage{booktabs, array}
\usepackage{tikz}
\usepackage{multirow}
\usepackage{pstool}
\usepackage{psfrag}
\usepackage{pictex}
\usepackage{pifont}
\usepackage{xcolor}
\usepackage{bm}
\usepackage{algorithm}
\usepackage{algorithmic}
\usetikzlibrary{shapes}
\usetikzlibrary{shapes.misc}
\usepackage{float}
\usepackage{amssymb}
\usepackage{newtxtext}
\usepackage{newtxmath}
\usepackage{natbib}
\usepackage{hyperref}
\hypersetup{
    colorlinks = true,
    urlcolor   = blue,
    citecolor  = black,
}

\newcommand{\RomanNumeralCaps}[1]
\linenumbers

\title{Optimal sensor and actuator placement for feedback control of vortex shedding}

\author{Bo Jin
  \corresp{\email{bjin1@student.unimelb.edu.au}},
  Simon J.  Illingworth
 \and Richard D. Sandberg}

\affiliation{Department of Mechanical Engineering, University of Melbourne, VIC, 3010, Australia}

\DeclarePairedDelimiter\norm{\lVert}{\rVert}

\newcommand{\blacklineshort}{\raisebox{2pt}{\tikz{\draw[black,solid,line width = 1pt](0,0) -- (1.5mm,0);}}}

\newcommand{\blacksmalldot}{\raisebox{1.5pt}{\tikz{\draw[black,fill=black] (0,0) circle (.2ex);}}}

\newcommand{\bluedot}{\tikz{\draw[blue,fill=blue] (0,0) circle (.4ex)}}

\newcommand{\mytriangle}[1]{\tikz{\node[draw=#1,fill=#1,isosceles triangle,isosceles triangle stretches,shape border rotate=90,minimum width=0.175cm,minimum height=0.175cm,inner sep=0pt] at (0,0) {};}}

\newcommand{\mydtriangle}[1]{\tikz{\node[draw=#1,fill=#1,isosceles triangle,isosceles triangle stretches,shape border rotate=-90,minimum width=0.175cm,minimum height=0.175cm,inner sep=0pt] at (0,0) {};}}

\newcommand{\mysquare}[1]{\tikz{\filldraw[draw=#1,fill=#1] (0,0) rectangle (0.15cm,0.15cm);}}

\newcommand{\dottedline}{\raisebox{2pt}{\tikz{\draw[black,dotted,line width = 1.25pt](0,0) -- (5mm,0);}}}

\newcommand{\squareline}[1]{\tikz{\filldraw[draw=#1,fill=white,line width = 1pt] (0,0) rectangle (0.175cm,0.175cm);\draw[#1,solid,line width = 1pt](-0.175cm,0.0875cm) -- (0.35cm,0.0875cm)}}

\newcommand{\cirline}[1]{\tikz{\filldraw[draw=#1,fill=white,line width = 1pt] (0,0) circle (0.1cm);\draw[#1,solid,line width = 1pt](-0.2625cm,0.0cm) -- (0.2625cm,0.0cm)}}

\newcommand{\triangleline}[1]{\tikz{\node[draw=#1,fill=white,isosceles triangle,isosceles triangle stretches,shape border rotate=90,minimum width=0.175cm,minimum height=0.175cm,inner sep=0pt,line width = 1pt] at (0,0) {};\draw[#1,solid,line width = 1pt](-0.2625cm,0.00625cm) -- (0.2625cm,0.00625cm)}}

\newcommand{\dtriangleline}[1]{\tikz{\node[draw=#1,fill=white,isosceles triangle,isosceles triangle stretches,shape border rotate=-90,minimum width=0.175cm,minimum height=0.175cm,inner sep=0pt,line width = 1pt] at (0,0) {};\draw[#1,solid,line width = 1pt](-0.2625cm,-0.01cm) -- (0.2625cm,-0.01cm)}}

\newcommand{\diamondfill}[1]{\tikz{\node[draw=#1,fill=#1,diamond,scale=0.6,xscale=0.8,line width = 1pt] at (0,0) {};}}

\tikzstyle{block} = [draw, fill=white!20, rectangle,minimum height=1em, minimum width=2em]
\tikzstyle{align} = [fill=white, rectangle, minimum height=0.1em, minimum width=0.1em]
\tikzstyle{sum} = [draw, fill=white!20, circle, inner sep=1pt,node distance=0.75cm]
\tikzstyle{input} = [coordinate]
\tikzstyle{output} = [coordinate]

\begin{document}
\maketitle
\begin{abstract}
We consider linear feedback control of the two-dimensional flow past a cylinder at low Reynolds numbers, with a particular focus on the optimal placement of a single sensor and a single actuator.
To accommodate the high dimensionality of the flow we compute its leading resolvent forcing and response modes to enable the design of $\mathcal{H}_2$-optimal estimators and controllers.
We then investigate three control problems: i) optimal estimation (OE) in which we measure the flow at a single location and estimate the entire flow; ii) full-state information control (FIC) in which we measure the entire flow but actuate at only one location; and iii) the overall feedback control problem in which a single sensor is available for measurement and a single actuator is available for control.
We characterize the performance of these control arrangements over a range of sensor and actuator placements and discuss implications for effective feedback control when using a single sensor and a single actuator.
The optimal sensor and actuator placements found for the OE and FIC problems are also compared to those found for the overall feedback control problem over a range of Reynolds numbers.
This comparison reveals the key factors and conflicting trade-offs that limit feedback control performance.
\end{abstract}

\section{Introduction}
Flow control is either passive without power input or active with powered actuators, which can be of tremendous benefit in a number of applications \citep{gad2007flow}. Typical examples include the altering and suppression of vortex shedding, enhanced mixing, drag reduction and noise abatement \citep{choi2008control,tan2018review,ceccio2010friction,gad2003flow}. In the last few decades, efforts have been made to improve the ability of manipulating fluid dynamics. One of the foremost challenges in flow control is the placement of control devices, which is crucial to the performance of both passive control \citep{strykowski1990formation,hwang2006control} and active control \citep{belson2013feedback}. Finding the optimal placement for control devices, although challenging, could significantly improve the effectiveness and efficiency of flow control schemes and provide deeper insights into the physical properties of fluid flows.

\subsection{Placement of control devices}
Most studies concerning optimal placement are based on the physical characteristics of flow systems \citep{schmid2014analysis,Chomaz05}. Some recent studies have suggested that any sensors should be placed where any unstable eigenmodes are large for the best detectability for those modes, and that any actuators should be placed where the corresponding adjoint modes are large for the best stabilisability for those adjoint modes \citep{lauga2003decay,aakervik2007optimal,bagheri2009input}. Similar arguments were also given based on a Gramian-based analysis of the open-loop system with full sensing and actuation: good sensor locations overlap with regions that have the largest response to external disturbances, as indicated by the leading eigenmodes of the controllability Gramian; good actuator locations overlap with regions that have the highest receptivity to perturbations, as indicated by the leading eigenmodes of the observability Gramian \citep{ma2011reduced,chen2014fluid}. 

Moreover, \cite{lauga2004performance} considered linear $\mathcal{H}_{\infty}$ feedback control of the complex Ginzburg–Landau (CGL) equation and placed an actuator and a sensor in the wavemaker region, which was originally introduced for the case of weakly-non-parallel flows \citep{chomaz1991frequency,monkewitz1993global}. Further work conducted by \cite{giannetti2007structural} defined a wavemaker region using an eigenvalue sensitivity analysis for strongly-non-parallel flows, e.g.~the cylinder flow, based on the concept of localised feedback of flow perturbations. Specifically, these regions describe the overlap between any unstable eigenmodes and their corresponding adjoint-eigenmodes, inside which local feedback mechanisms could result in large modifications of any unstable eigenvalues to push them into the stable half-plane. Based on this eigenvalue sensitivity analysis, \cite{camarri2010feedback} determined the types and positions of sensors as well as feedback coefficients for a simple proportional feedback control law for the flow past a square cylinder confined in a channel. Their strategy led to the successful stabilisation of the flow up to a Reynolds number that was 100$\%$ higher than the critical value after which otherwise the flow would become unsteady. Similar studies have employed various sensitivity analyses for the selection and placement of collocated actuator–sensor pairs in a separated boundary layer \citep{natarajan2016actuator} and optimal sensor placement for variational data assimilation of unsteady flows \citep{mons2017optimal}. Rather than control perturbations, \cite{marquet2008sensitivity} assumed a steady forcing for base-flow modifications and reproduced flow-stabilising regions using sensitivity analysis, which showed good agreement with those found experimentally by \cite{strykowski1990formation}. A more detailed comparison between these studies is presented in \cite{sipp2010dynamics}.

Although the modal analyses described above provide sensible placements for control purposes, they do not yield the true optimal placement due to the strong non-normality of fluid flows. Therefore, a rigorous methodology and justification for finding the optimal placement should be based on the optimal control performance of each possible sensor-actuator configuration. Standard metrics of quantifying control performance include the $\mathcal{H}_2$ norm which measures the energy of the system's impulse response and the $\mathcal{H}_{\infty}$ norm which measures the worst-case (i.e.~most amplified) response to harmonic forcing \citep[pp.~368-382]{skogestad2007multivariable}. However, it is challenging to recompute the optimal control performance for each new placement of sensors and actuators, especially for high-dimensional control problems arising from two- or three-dimensional flows. \cite{lauga2003decay} considered the one-dimensional Ginzburg-Landau system for which a full-state information controller (i.e.~linear quadratic regulator) was designed for each possible actuator position. The variation of the optimal actuator location with Reynolds number was compared to that predicted by eigenanalysis. Recent studies have also considered optimal sensor placement for state estimation of the one-dimensional dispersive wave equation \citep{khan2015computation} and the Boussinesq equations that model a controlled thermal fluid \citep{hu2016sensor}. Reduced-order modelling has also been employed in optimal placement problems for two-dimensional flows, such as the flow over a backward-facing step \citep{juillet2013control} and the cylinder flow \citep{akhtar2015using,jin2020feedback}.

There are only few studies that rigorously analyse the optimal placements of sensors and actuators and their implications for the flow's closed-loop dynamics. \cite{chen2011h,chen2014fluid} studied the optimal placement problem for $\mathcal{H}_2$ optimal control of the complex Ginzburg-Landau equation. They found the optimal sensor and actuator positions using an extended gradient minimisation algorithm developed by \cite{hiramoto2000optimal} for closed-loop control setups. In particular, the optimal placements of a single sensor and a single actuator were compared to those predicted by modal analyses, which demonstrated the shortcomings of eigenmode analysis and Gramian-based analysis for predicting optimal placements. \cite{chen2011h} comment that these shortcomings are caused either by the strong non-normality of the system characterised by non-orthogonal eigenmodes or by the presence of time delays. They further comment that the wavemaker region proposed by \cite{giannetti2007structural}, which indicates areas of high dynamical sensitivity, provides improved estimates of the optimal actuator and sensor placements. \cite{oehler2018sensor} further investigated optimal actuator and sensor placements for the same system but over a wider range of stability parameters. Their results indicated that the wavemaker region has no special significance for optimal placement and that, with increasing instability, the optimal placements move further away from that predicted by the wavemaker region. Instead, the optimal sensor and actuator positions show good agreement with those computed from the optimal-estimation and full-state-information control problems. A recent study of \cite{jin2020feedback} investigated optimal sensor placement in the two-dimensional cylinder wake using resolvent-based model-order reduction. A fundamental trade-off was demonstrated between measuring downstream information and reducing the time lag with respect to the actuator upstream. However, it is still not well understood whether feedback control performance is limited predominantly by the measurements (e.g.~sensor placement) or by the actuation (e.g.~actuator placement) or by their interaction in the overall feedback loop.

\subsection{Objectives of the present work}
The current work focuses on the optimal sensor and actuator placements for feedback control of the two-dimensional cylinder flow and investigates any trade-offs and coupling effects in the optimal placements. In particular, we first consider three optimal placement problems: i) the optimal estimation (OE) problem in which the objective is to estimate the entire flow using a single sensor; ii) the full-state information control (FIC) problem in which the entire flow field is known but only a single actuator is available for control; and iii) the collocated input-output control (CIOC) problem in which a single sensor is available for measurements, which is collocated with a single actuator for control (localised feedback). By varying the Reynolds number and therefore the stability of the flow, any fundamental limitations or trade-offs are made clear for the optimal placements of a single sensor (OE), of a single actuator (FIC) and for localised feedback (CIOC). 

The optimal performance achieved in the above three problems are compared to those achieved in the overall feedback control problem where a single sensor and a single actuator are separately placed at i) the optimal positions found for the OE and FIC problems, respectively; and ii) the optimal positions found for the overall feedback control problem. This provides a benchmark for evaluating the extent to which the optimal placements for the OE, FIC and CIOC problems approximate the optimal feedback control setup and reveals any key factors that limit control performance. We discuss implications for sensor placement, actuator placement and the coupling effect between sensing and actuation (i.e.~the time lag) for effective feedback control using a single sensor and a single actuator.

The work is organised as follows. Mathematical formulations and flow configurations are given in \S\ref{sec:math_formulation}. Arrangements and design method of the estimation and control problems are introduced in \S\ref{sec:method}. In \S\ref{sec:results}, we present results and discussions about the optimal sensor and actuator placements for feedback control of the two-dimensional cylinder flow. Conclusions are drawn in \S\ref{sec:conclusions}.

\section{Mathematical formulation}\label{sec:math_formulation}
\subsection{Governing equations}
We consider the incompressible flow past a two-dimensional circular cylinder. The incompressible Navier-Stokes equations describe the conservation of mass and momentum of an incompressible fluid:
\begin{equation}\label{equ:nsequation}
    \begin{gathered}
        \partial_t\textbf{\textit{u}}=-\textbf{\textit{u}}\cdot\nabla\textbf{\textit{u}}-\nabla \textit{p}+\Rey^{-1} \nabla^2\textbf{\textit{u}}\ ,\\
        \nabla\cdot\textbf{\textit{u}}=0\ ,
    \end{gathered}
\end{equation}
where the Reynolds number $\Rey={\textit{U}_{\infty}D}/{\nu}$ is based on a uniform inflow velocity $\textit{U}_{\infty}$ and the cylinder diameter $D$ to make all variables dimensionless. Here, $\nu$ is the kinematic viscosity. In this study, we focus on Reynolds numbers in the range $\Rey \in [50,\ 110]$ for which the cylinder wake has a single linearly unstable mode that drives the flow to periodic self-sustained limit-cycle oscillations (vortex shedding). The objective of feedback control is to completely suppress vortex shedding behind a two-dimensional circular cylinder and drive the flow towards its unstable steady state (base flow). Therefore, we linearise the nonlinear Navier-Stokes equations \eqref{equ:nsequation} about the laminar base flow ($\textbf{\textit{U}},\ \textit{P}$) which allows us to use existing linear control theory and analysis techniques:
\begin{equation}\label{equ:pertur}
  \begin{gathered}
    \partial_t\textbf{\textit{u}}'=\mathcal{L}\textbf{\textit{u}}'-\nabla\textit{p}'+\textbf{\textit{f}}'\ ,\\
    \nabla\cdot\textbf{\textit{u}}'=0.
  \end{gathered}
\end{equation}

The perturbations ($\textbf{\textit{u}}',\ \textit{p}'$) evolve according to the linear operator $\mathcal{L} = {-\textbf{\textit{U}}\cdot\nabla()} - {()\cdot\nabla\textbf{\textit{U}}} + {\Rey^{-1}\nabla^2()}$. The remaining nonlinear terms $ {-\textbf{\textit{u}}'\cdot\nabla\textbf{\textit{u}}'}$ are neglected due to the assumption of small perturbations and the source term $\textbf{\textit{f}}'$ models any external forcing, such as stochastic disturbances or actuation. The linear perturbation equations \eqref{equ:pertur} can also be written compactly as
\begin{equation}\label{equ:compact_form}
    \dfrac{\partial}{\partial t}
    \underbrace{
    \begin{bmatrix}
        \text{I}&\text{0}\\
        \text{0}&\text{0}
    \end{bmatrix}}_{\mathcal{E}}
    \begin{bmatrix}
        \textbf{\textit{u}}'\\
        \textit{p}'
    \end{bmatrix}
    =
    \underbrace{
    \begin{bmatrix}
    \mathcal{L}&-\nabla()\\
    -\nabla\cdot()&\text{0}
    \end{bmatrix}}_{\mathcal{A}}
    \begin{bmatrix}
        \textbf{\textit{u}}'\\
        \textit{p}'
    \end{bmatrix}+
    \underbrace{
    \begin{bmatrix}
        \text{I}\\
        \text{0}
    \end{bmatrix}}_{\mathcal{P}}\textbf{\textit{f}}'\ ,
\end{equation}
where $\mathcal{E}=\mathcal{P}\mathcal{P}^T$ and $\mathcal{P}$ is the prolongation operator that maps a velocity vector $\textbf{\textit{u}}'$ to a velocity-zero-pressure vector $[\textbf{\textit{u}}',\ \text{0}]^T$. $\mathcal{A}$ denotes the linearized Navier–Stokes operator around the base flow.

\subsection{Flow configuration and discretisation}\label{sec:numerical_setups}
\begin{figure}
  \centerline{\includegraphics[width=0.8\textwidth]{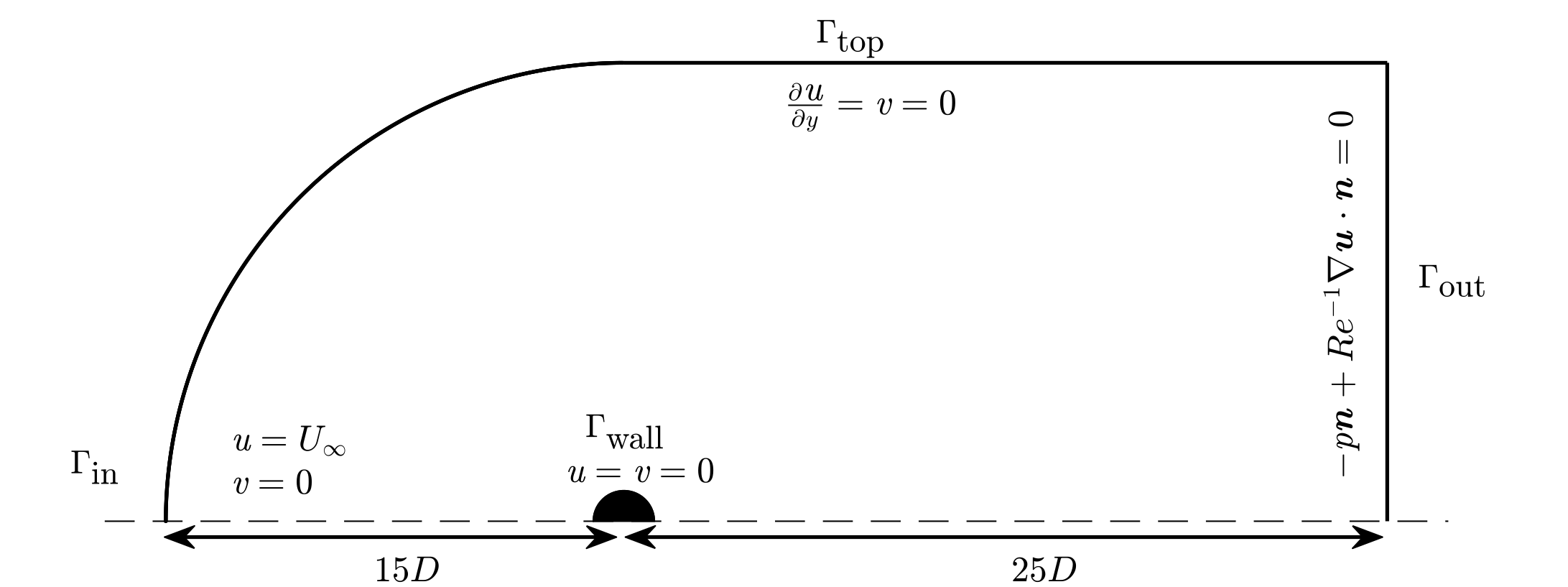}}
  \caption[Computational domain and boundary conditions]{Computational domain and boundary conditions for the cylinder flow (only a one-half segment of the entire geometry is shown).}
\label{fig:compudomain}
\end{figure}

The schematic diagram of the setup used for the two-dimensional cylinder flow is displayed in figure \ref{fig:compudomain}, in which we employ the same computational domain and boundary conditions as those used by \cite{leontini2006wake,jin2020feedback}. A uniform freestream velocity ($\textit{U}_{\infty}=1$) is imposed at the inlet boundary ($\Gamma_{in}$) and encounters a cylinder ($\Gamma_{wall}$) of diameter $D=1$ with no-slip boundary conditions. Symmetric boundary conditions and standard outflow boundary conditions are imposed at the top boundary ($\Gamma_{top}$) and the outlet boundary ($\Gamma_{out}$), respectively. Note that the linear perturbation system has the same boundary conditions as those depicted in the figure except at the inlet where homogeneous boundary conditions ($\textbf{\textit{u}}'=[0,\ 0]$) are enforced to ensure zero perturbations at infinity. 

The Navier-Stokes equations are discretized using Taylor-Hood finite elements over a structured mesh using the FEniCS platform \citep{logg2012automated}. The mesh points are clustered smoothly near the cylinder and in the wake to appropriately resolve the details of the flow. In particular, the mesh consists of $2.7\times 10^4$ triangles and the minimum wall-normal spacing around the cylinder is 0.01. The compound state vector $\textbf{\textit{w}}=[\textbf{\textit{u}}',\ \textit{p}']^T\in\mathbb{R}^{N}$ thus has over $N=1.2\times 10^5$ degrees of freedom. The laminar base flow governed by the steady Navier–Stokes equations is then solved for using a Newton method. We use a backward Euler scheme for time discretization ($\Delta t=0.01$) in numerical simulations of the linear perturbation system. Note that the laminar base flows and discretised perturbation systems have been validated by comparing them with the stability analysis results of \cite{Barkley06}. A sparse direct LU solver (MUMPS, \cite{amestoy2001fully}) and iterative Arnoldi methods (ARPACK, \cite{lehoucq1998arpack}) are used for all linear problems encountered in the study. 
\section{Modelling and control methods}\label{sec:method}
We now consider the feedback control of linear perturbations in the flow. This section is organised as follows. The discretised input-output system is formulated in \S\ref{sec:i-o_system}. The estimation and control arrangements are then presented in \S\ref{sec:control_setups}. In \S\ref{sec:design_method}, we introduce a resolvent-based design approach for the estimation and control problems. 

\subsection{Input-output system}\label{sec:i-o_system}
\begin{figure}
  \centerline{\includegraphics[width=0.6\textwidth]{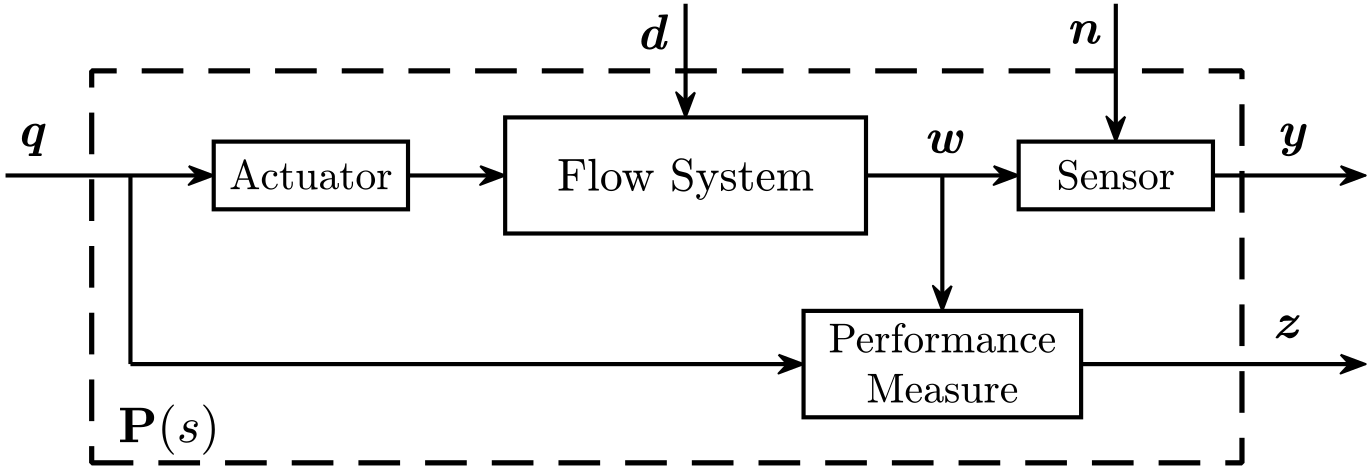}}
  \caption[Block diagram of the state-space model.]{Block diagram of the state-space model \eqref{equ:ssmodel_summary}, which is denoted as a transfer function $\textbf{P}(s)$ with the Laplace variable $s$.}
\label{fig:blockdiagram}
\end{figure}

The linear system \eqref{equ:compact_form} is subject to stochastic disturbances and actuation, which serve as perturbations and control mechanisms for the flow. Following spatial discretization, we can express \eqref{equ:compact_form} as a linear time-invariant state-space model $\textbf{P}(s)$ with outputs of interest (i.e.~$\textbf{\textit{y}}$ and $\textbf{\textit{z}}$), as depicted in figure \ref{fig:blockdiagram}:
\begin{gather}\label{equ:ssmodel_summary}
    \begin{aligned}
        \textbf{E}\Dot{\textbf{\textit{w}}}&=\textbf{A}\textbf{\textit{w}}+\textbf{B}_q\textbf{\textit{q}}+\textbf{B}_d\textbf{\textit{d}}\\
        \textbf{\textit{y}}&=\textbf{C}_y\textbf{\textit{w}}+\textbf{V}^{1/2}\textbf{\textit{n}}\\
        \textbf{\textit{z}}&=[\textbf{C}_z\textbf{\textit{w}}\hspace{3.5mm} \textbf{R}^{1/2}\textbf{\textit{q}}]^T\ ,
    \end{aligned}
\end{gather}
where the compound state vector $\textbf{\textit{w}}=[\textbf{\textit{u}}',\ \textit{p}']^T$ and the matrix $\textbf{E}=\textbf{P}\textbf{M}\textbf{P}^T$. Here, $\textbf{P}$ and $\textbf{M}$ denote the prolongation matrix and the mass matrix of the velocity state due to the spatial discretization. The spatial discretization of actuation and statistical properties of disturbances are represented by the matrices $\textbf{B}_q$ and $\textbf{B}_d$, respectively. The disturbances are modelled as uncorrelated zero-mean Gaussian
white noise and are injected over the entire velocity field. Therefore, the statistical properties of disturbances is given by $\textbf{B}_d=\textbf{P}\textbf{M}^{1/2}$ after the spatial discretization \citep{croci2018efficient}. 

The sensor measurement $\textit{\textbf{y}}$ provides sensing and is characterised by the matrix $\textbf{C}_y$. It includes a contribution from sensor noise \textbf{\textit{n}} which is white in space and time with magnitude $\alpha$ (i.e.~$\textbf{V}^{1/2}=\alpha\textbf{I}$). We aim to minimise the mean kinetic energy of linear perturbations $\textbf{\textit{w}}$. That is, the $\mathcal{H}_2$ norm of the performance measure $\textbf{\textit{z}}$ is minimised (with $\textbf{C}_z=\textbf{M}^{1/2}\textbf{P}^T$). Therefore the cost function is of the form:
\begin{equation}\label{equ:cost_j}
    \textbf{\textit{J}}=\lim_{t\rightarrow\infty}\dfrac{1}{T}\int_0^T\textbf{\textit{z}}^T\textbf{\textit{z}}\ dt\ .
\end{equation}
Note that the actuation input $\textbf{\textit{q}}$ is a signal of interest in the control problem, which is scaled by $\beta$ to ensure that the control effort is sensible (i.e.~$\textbf{R}^{1/2}=\beta\textbf{I}$). 

\subsection{Estimation and control setups}\label{sec:control_setups}
\newcolumntype{a}{>{\centering\arraybackslash}m{0.47\textwidth}}
\newcolumntype{b}{>{\centering\arraybackslash}m{0.94\textwidth}}
\aboverulesep=0ex
\belowrulesep=0ex
\begin{figure*}
\begin{tabular}{@{}|a|a|@{}}
\toprule
\multicolumn{2}{|b|}{
    \hspace{0mm}
	\centerline{
	\includegraphics[width=0.94\textwidth]{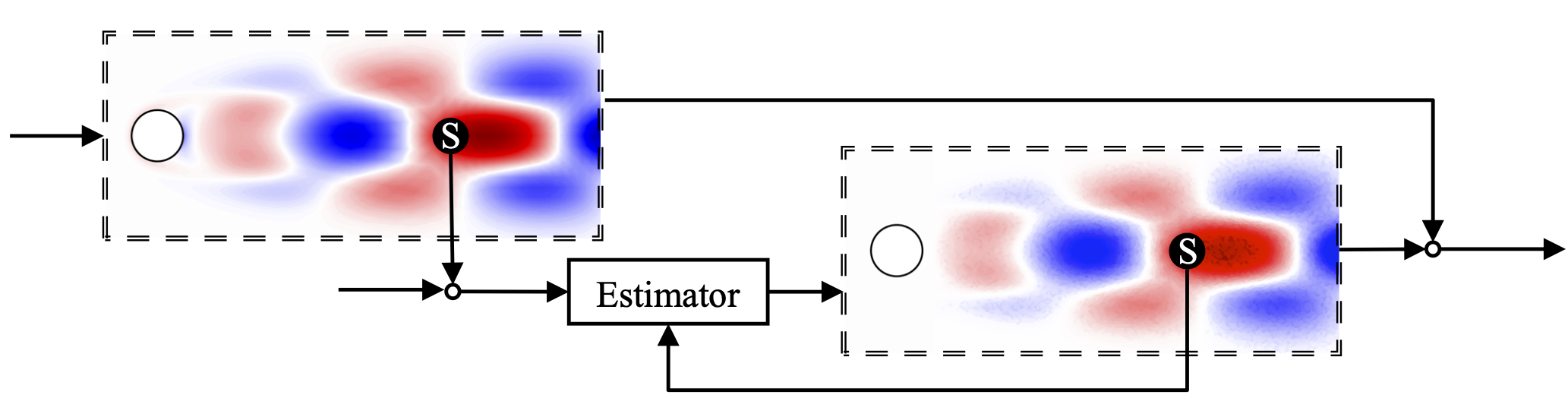} 
	\llap{\parbox[b]{5.0in}{(a) OE \\\rule{0ex}{1.35in}}}
    \llap{\parbox[b]{3.475in}{\textit{\textbf{y}}\\\rule{0ex}{0.27in}}}
    \llap{\parbox[b]{3.1in}{$\textit{\textbf{y}}_e$\\\rule{0ex}{0.075in}}}
    \llap{\parbox[b]{2.35in}{$\textit{\textbf{w}}_e$\\\rule{0ex}{0.7in}}}
    \llap{\parbox[b]{4.75in}{$\textit{\textbf{w}}$\\\rule{0ex}{1.1in}}}
    \llap{\parbox[b]{0.475in}{$\textit{\textbf{e}}$\\\rule{0ex}{0.545in}}}
    \llap{\parbox[b]{5.075in}{$\textit{\textbf{d}}$\\\rule{0ex}{0.915in}}}
    \llap{\parbox[b]{4.05in}{$\textit{\textbf{n}}$\\\rule{0ex}{0.29in}}}
	}
} \\[0pt] \midrule
velocity sensor at $\textbf{\textit{x}}_s$ & no actuator \\ \midrule
\multicolumn{2}{b}{}\\\midrule
\multicolumn{2}{|b|}{
    \vspace{2mm}
	\centerline{
	\hspace{7.5mm}
	\includegraphics[width=0.94\textwidth]{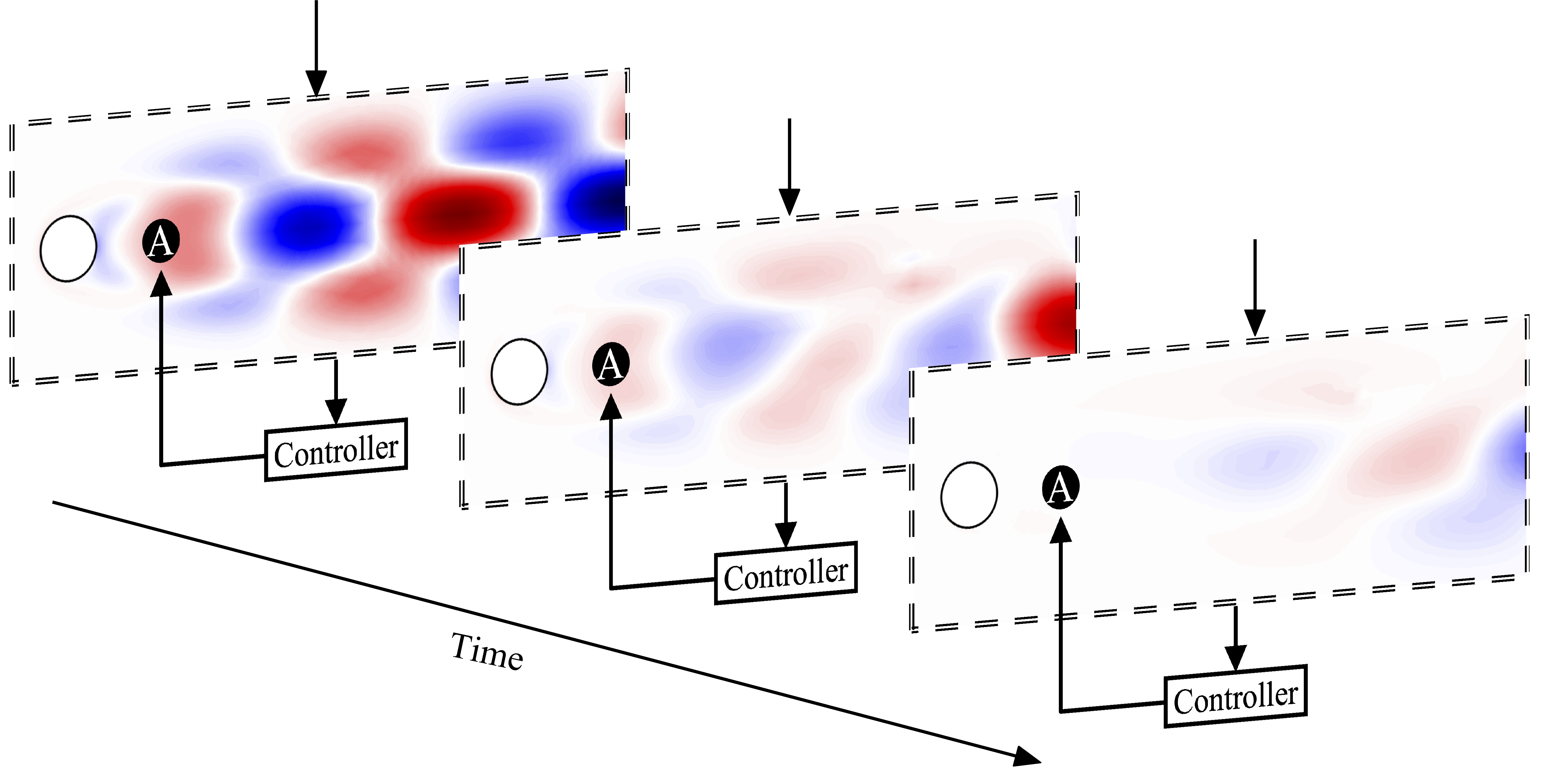} 
	\llap{\parbox[b]{5.125in}{(b) FIC\\\rule{0ex}{2.35in}}}
    \llap{\parbox[b]{4.95in}{$\textit{\textbf{w}}(t_0)$\\\rule{0ex}{1.95in}}}
    \llap{\parbox[b]{4.675in}{\textit{\textbf{q}}\\\rule{0ex}{1.1in}}}
    \llap{\parbox[b]{4.225in}{\textit{\textbf{d}}\\\rule{0ex}{2.32in}}}
    \llap{\parbox[b]{3.575in}{$\textit{\textbf{w}}(t_1)$\\\rule{0ex}{1.55in}}}
    \llap{\parbox[b]{3.335in}{\textit{\textbf{q}}\\\rule{0ex}{0.725in}}}
    \llap{\parbox[b]{2.8in}{\textit{\textbf{d}}\\\rule{0ex}{1.925in}}}
    \llap{\parbox[b]{2.225in}{$\textit{\textbf{w}}(t_2)$\\\rule{0ex}{1.15in}}}
    \llap{\parbox[b]{1.975in}{\textit{\textbf{q}}\\\rule{0ex}{0.325in}}}
    \llap{\parbox[b]{1.385in}{\textit{\textbf{d}}\\\rule{0ex}{1.55in}}}
	}
} \\[-10pt] \midrule
no sensor & body force actuator at $\textbf{\textit{x}}_a$ \\ \midrule
\multicolumn{2}{b}{}\\\midrule
\multicolumn{2}{|b|}{
    \vspace{2mm}
	\centerline{
	\hspace{12.5mm}
    \includegraphics[width=0.94\textwidth]{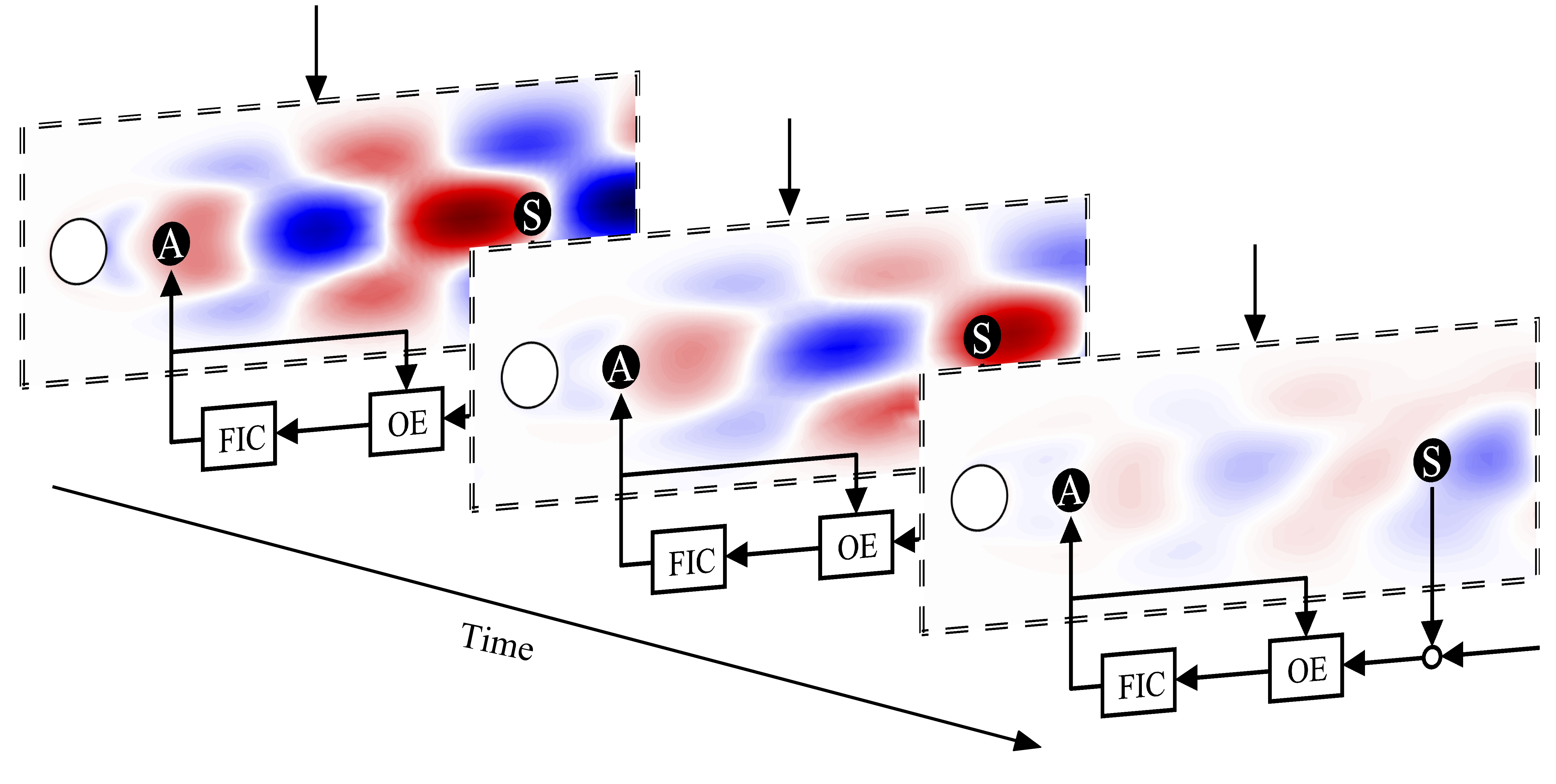}
    \llap{\parbox[b]{5.15in}{(c) IOC / CIOC\\\rule{0ex}{2.35in}}}
    \llap{\parbox[b]{0.275in}{\textit{\textbf{n}}\\\rule{0ex}{0.295in}}}
    \llap{\parbox[b]{0.625in}{\textit{\textbf{y}}\\\rule{0ex}{0.25in}}}
    \llap{\parbox[b]{1.775in}{\textit{\textbf{q}}\\\rule{0ex}{0.365in}}}
    \llap{\parbox[b]{1.275in}{$\textit{\textbf{w}}_e$\\\rule{0ex}{0.2in}}}
    \llap{\parbox[b]{1.275in}{\textit{\textbf{d}}\\\rule{0ex}{1.55in}}}
    
    \llap{\parbox[b]{3.325in}{\textit{\textbf{q}}\\\rule{0ex}{0.75in}}}
    \llap{\parbox[b]{2.775in}{$\textit{\textbf{w}}_e$\\\rule{0ex}{0.605in}}}
    \llap{\parbox[b]{2.865in}{\textit{\textbf{d}}\\\rule{0ex}{1.925in}}}
    
    \llap{\parbox[b]{4.855in}{\textit{\textbf{q}}\\\rule{0ex}{1.15in}}}
    \llap{\parbox[b]{4.325in}{$\textit{\textbf{w}}_e$\\\rule{0ex}{1.00in}}}
    \llap{\parbox[b]{4.475in}{\textit{\textbf{d}}\\\rule{0ex}{2.285in}}}
    
    \llap{\parbox[b]{5.25in}{$\textit{\textbf{w}}(t_0)$\\\rule{0ex}{1.95in}}}
    \llap{\parbox[b]{3.85in}{$\textit{\textbf{w}}(t_1)$\\\rule{0ex}{1.55in}}}
    \llap{\parbox[b]{2.45in}{$\textit{\textbf{w}}(t_2)$\\\rule{0ex}{1.15in}}}
    }
} \\[-10pt] \midrule
velocity sensor at $\textbf{\textit{x}}_s$ & body force actuator at $\textbf{\textit{x}}_a$ \\ \bottomrule
\end{tabular}
\caption[Setups of the OE, FIC and IOC problems]{Setups of the OE, FIC and IOC problems. (a) Optimal estimation (OE) of the whole flow field using a single sensor at $\textbf{\textit{x}}_s$. (b) Full-state information control (FIC) using a single actuator at $\textbf{\textit{x}}_a$ when the entire flow field is known. (c) Input-Output feedback control (IOC) with a single sensor at $\textbf{\textit{x}}_s$ and a single actuator at $\textbf{\textit{x}}_a$.} \label{fig:controlsetups}
\end{figure*}

We employ $\mathcal{H}_2$-optimal control tools as established by \cite{doyle1988state} to solve the feedback control (i.e.~input-output control) problem for the linear system \eqref{equ:ssmodel_summary}. A complete introduction to the method can be found in \cite{skogestad2007multivariable}, in which any input-output control problem is composed of two basic problems: i) an optimal estimation problem and ii) a full-state information control problem. Figure \ref{fig:controlsetups} shows the three setups for the estimation and control problems considered: 
\begin{enumerate}
    \item \textit{the optimal estimation (OE) problem}, where the entire flow field is estimated using a single sensor that measures the perturbation velocity $\textbf{\textit{u}}'(\textbf{\textit{x}}_s,t)$ at a single point in the flow. The sensor is contaminated by noise $\textbf{\textit{n}}$ of magnitude $\alpha=10^{-4}$ so that sensor noise is present but minimal. For each sensor placement, we aim to minimise the mean (i.e.~time-averaged) kinetic energy of the estimation error ($\textbf{\textit{e}}=\textbf{\textit{w}}-\textbf{\textit{w}}_e$) under the excitation of stochastic disturbances and in the presence of sensor noise. The optimal sensor position therefore leads to the smallest possible $\mathcal{H}_2$ norm of the estimation error (i.e.~the optimal estimation of the entire cylinder flow). 
    
    \item \textit{the full-state information control (FIC) problem}, where the entire flow field is known (i.e.~measured perfectly everywhere without any sensor noise $\alpha=0$) but only a single body force $\textbf{\textit{f}}'(\textbf{\textit{x}}_a,t)$ that serves as an in-flow actuator is available for control and operates according to the signal $\textbf{\textit{q}}$. For each actuator placement, the task of the FIC problem is to use a sensible control effort to minimise the mean kinetic energy of flow perturbations that are excited by stochastic disturbances. We choose a small control penalty of $\beta=10^{-4}$ to allow for relatively aggressive control. Analogous to the OE problem, the optimal actuator position achieves the smallest possible $\mathcal{H}_2$ norm of the perturbation velocity (i.e.~the optimal FI control of the entire cylinder flow). 
    
    \item \textit{the collocated input-output control (CIOC) problem}, where a single collocated actuator-sensor pair is available for both control and measurement (i.e.~$\textbf{\textit{x}}_s=\textbf{\textit{x}}_a$). In this case, we use the same setups as those used for the OE and FIC problems (i.e.~$\alpha=\beta=10^{-4}$) but the single sensor and single actuator are collocated to model a localised feedback control mechanism with minimal time lag. For each collocated actuator-sensor placement, we aim to minimise the mean kinetic energy of flow perturbations under excitation from external disturbances and in the presence of sensor noise. The optimal position for the collocated actuator-sensor pair thus provides the best compromise between adequate estimation of the entire flow and adequate FI control of the entire flow.
    
    \item \textit{the general input-output control (IOC) problem}, which shares the same setup as that described in the CIOC problem but which uses a single sensor placed downstream for measurement and a single actuator placed upstream for control (i.e.~$\textbf{\textit{x}}_s\neq \textbf{\textit{x}}_a$). In particular, we consider the sensor and actuator placements at i) the optimal positions that achieve the best feedback control performance; and ii) the optimal locations found for the OE and FIC problems, respectively. In the latter case, the sensor and actuator placements provide the best estimation performance and the best FI control performance, respectively, of the whole cylinder flow but may allow excessive time lag between the sensor and the actuator. By comparing it to the above three problems, we aim to evaluate the coupling effect between sensing and actuating, e.g.~the time-lag effect, for effective feedback control.
\end{enumerate}

Having defined the estimation and control setups, we then need to solve the OE and FIC problems (e.g.~solve their corresponding Riccati equations). Based on the Separation Theorem \citep{georgiou2013separation}, any IOC problem can be solved by combining the solutions of the corresponding OE and FIC problems. Refer to Appendix \ref{sec:app.a} for a summary of the systems and solutions for these problems. However, the generalised algebraic Riccati equations associated with the OE and FIC problems are generally of high dimension. Although it is common to perform numerical simulations for two- or three-dimensional fluid flows (i.e.~$N>10^5$), traditional control design tools (e.g.~Riccati solvers) typically become computationally intractable for $N>10^3$. This difficulty has been partially overcome by a sparse Riccati solver using an extended low-rank method, in which the number of inputs and outputs (so-called terminals) is limited to be far less than the dimension of the control problem \citep{benner2019mess,SaaKB19-mmess-2.0}. For problems with either many inputs (e.g.~full-state disturbances in the OE problem) or many outputs (e.g.~full-state measuring in the FIC problem), no efficient numerical tools are available to directly handle large-scale systems. In the next section, we will introduce a `terminal reduction' method to overcome the challenges of many inputs and outputs.

\subsection{Optimal estimator and controller design}\label{sec:design_method}
\begin{figure}
\vspace{2mm}
  \centerline{\includegraphics[width=0.65\textwidth]{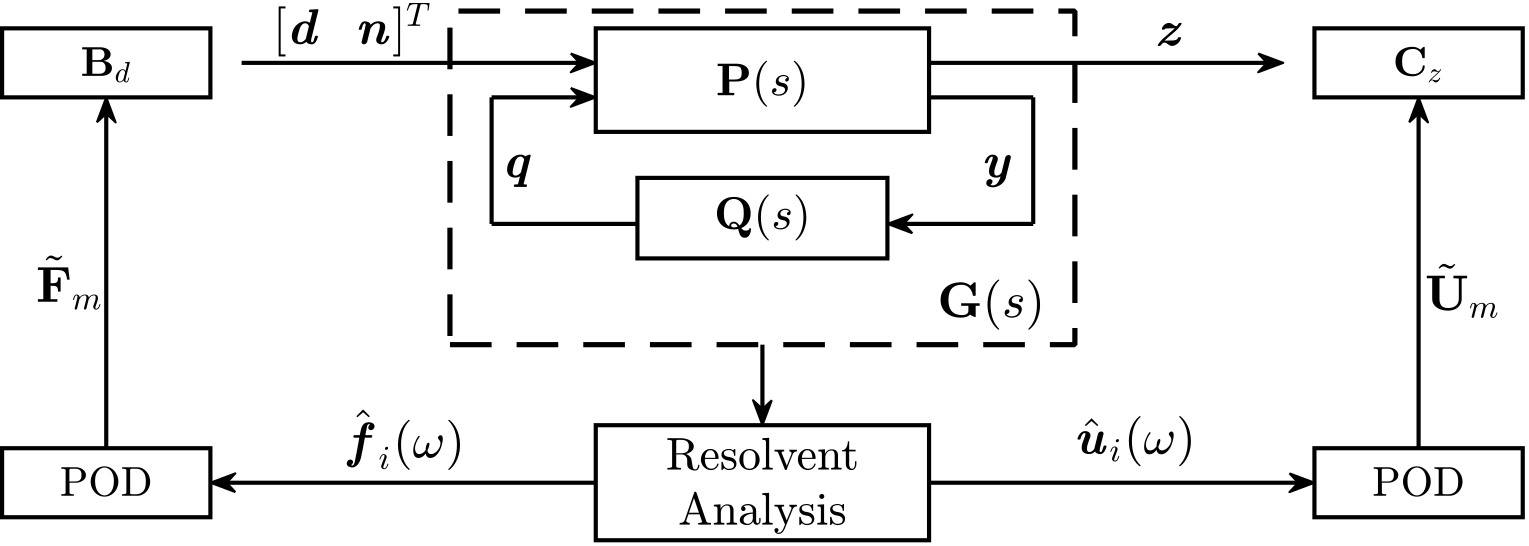}}
  \caption[Schematic of full-dimensional design procedure for the optimal estimation and control of a high-dimensional flow system.]{Schematic of the resolvent-based design approach for the optimal estimation and control of a high-dimensional flow system. The state-space model \eqref{equ:ssmodel_summary} is denoted as $\textbf{P}(s)$ and the closed-loop $\textbf{G}(s)$ is formed by coupling the secondary transfer function $\textbf{R}(s)$ from the sensor measurement $\textbf{\textit{y}}$ to the control signal $\textbf{\textit{q}}$.}
\label{fig:methoddiagram}
\end{figure}

As depicted in figure \ref{fig:methoddiagram}, the closed-loop transfer function $\textbf{G}(s)$ can be formed once the estimator or controller is designed, defined such that $\textbf{\textit{z}}=\textbf{G}(s)[\textbf{\textit{d}},\textbf{\textit{n}}]^T$. The feedback law from the sensor measurement $\textbf{\textit{y}}$ to the actuation signal $\textbf{\textit{q}}$ is represented by the transfer function $\textbf{Q}(s)$. For both the OE and FIC problems, our purpose is to minimise $\textbf{G}(s)$ such that the performance measurement $\textbf{\textit{z}}$ is small. Therefore, an equivalent form of the cost function \eqref{equ:cost_j} is the $\mathcal{H}_2$ norm of $\textbf{G}(s)$:
\begin{gather}\label{equ:J_h2norm}
    \begin{aligned}
        \textbf{\textit{J}}=\norm{\textbf{G}(s)}^2_2&=\dfrac{1}{2\pi}\int_{-\infty}^{\infty}\rm{tr}\{\textbf{G}^H(j\omega)\textbf{G}(j\omega)\}\ d\omega\\
         &=\dfrac{1}{2\pi}\int_{-\infty}^{\infty}\sum_{i}^{}\sigma^2_i(j\omega)\ d\omega\ ,
    \end{aligned}
\end{gather}
where $\sigma_i(j\omega)$ are the singular values of the transfer function $\textbf{G}(s)$ at frequency $\omega$ arranged in descending order. The singular values of a transfer function can be considered as energy gains between a series of inputs and the corresponding outputs. We thus aim to minimise the integrated energy gain \eqref{equ:J_h2norm} for inputs and outputs over all frequencies and all possible directions. 

However, it is not feasible to consider all inputs or outputs while designing estimators or controllers for a high-dimensional flow system. One possible solution is to consider an alternative cost function $\gamma^2{(k,\omega_n)}$ which only includes a limited number of singular values within a specified frequency range:
\begin{equation}\label{equ:h2norm_resovent}
     \gamma^2(k,\omega_n)=\dfrac{1}{2\pi}\int_{-\omega_n}^{\omega_n}\sum_{i=1}^{k}\sigma^2_i(j\omega)\ d\omega\ . 
\end{equation}
Therefore only the first $k$ orthogonal inputs and outputs across a limited frequency range $\omega\in[-\omega_n,\ \omega_n]$ will be considered. This cost function is constructed based on two insights: i) for fluid flows, only a limited number of dominant physical mechanisms occur within a finite frequency range, e.g.~the instability of the linearised cylinder flow occurs around $\omega_c\approx 0.8$; ii) these physical mechanisms can be approximated by a small number of orthogonal inputs and outputs that have large energy gains $\sigma_i^2$, which are also the most significant for estimation or control. Instead of minimising all energy gains over all frequencies and all possible directions, it is more feasible to use the alternative cost function \eqref{equ:h2norm_resovent} that considers a significantly smaller number of inputs and outputs. 

The implementation of the alternative cost function is depicted in figure \ref{fig:methoddiagram}, where we iteratively replace $\textbf{B}_d$ and $\textbf{C}_z$ with properly constructed low-rank matrices (i.e.~$\textbf{B}_d=\textbf{P}\textbf{M}\Tilde{\textbf{F}}_m$ and $\textbf{C}_z=\Tilde{\textbf{U}}_m\textbf{M}\textbf{P}^T$). In this case, the white noise disturbances applied everywhere are limited to orthonormal modes in the low-rank input basis $\Tilde{\textbf{F}}_m$. As for the full-state performance measure $\textbf{\textit{z}}$, the system states $\textbf{\textit{w}}$ are recast as linear combinations of orthonormal output modes in the low-rank output basis $\Tilde{\textbf{U}}_m$. Therefore, the number of either inputs or outputs is reduced to $m$ --- the rank of the input and output bases. The construction of low-rank bases is based on the proper orthogonal decomposition (POD) of the first $k$ resolvent modes across a wide frequency range $[-\omega_n,\ \omega_n]$, which generates orthonormal POD modes ranked by their importance (i.e.~the energy gain). In this study, the low-rank bases $\Tilde{\textbf{F}}_m$ and $\Tilde{\textbf{U}}_m$ contain the first $m$ POD modes such that all relevant resolvent input and output modes can be recovered from linear combinations of orthonormal POD modes within a relative mismatch less than $10^{-6}$. Note that resolvent analysis preferentially displays low-rank characteristics for physical mechanisms that are active in fluid flows \citep{mckeon2010critical,sipp2013characterization}. In other words, there often exists a large separation between singular values $\sigma_i$ such that only a limited number of forcing modes give rise to energetic responses that are the most important for estimation and control. By choosing a sufficient number of resolvent modes over a sufficiently large frequency range, the resulting performance should eventually converge to the true global optimum. In this study, we choose the parameter combination $k=3$ and $\omega_n=9$, which is sufficient to achieve convergence for both the optimal performance and the optimal placements (see Appendix \ref{sec:app.b}). 

\section{Results}\label{sec:results}
We now design optimal estimators and controllers for the two-dimensional cylinder flow and find the optimal sensor and actuator placements. This section consists of three parts: (i) the optimal estimation (OE) problem with a single sensor (figure \ref{fig:controlsetups}(a)); (ii) the full-state information control (FIC) problem with a single actuator (figure \ref{fig:controlsetups}(b)); (iii) the collocated input-output control (CIOC) problem with a single collocated actuator-sensor pair (figure \ref{fig:controlsetups}(c)). In the last subsection \S\ref{sec:further}, we further consider an input-output control (IOC) setup with a single sensor and a single actuator placed at the optimal positions found for the OE and FIC problems, respectively (figure \ref{fig:controlsetups}(d)). The comparison of the optimal performance among all four cases provides deeper insights into the sensor and actuator placement problems and the influence of their coupling for effective feedback control.

\subsection{Optimal estimation problem}\label{sec:lqe}
\subsubsection{Brute-force sampling}\label{sec:lqe_bfs}
\begin{figure}
    \centerline{
    \hspace{2mm}
    \includegraphics[width=0.975\textwidth]{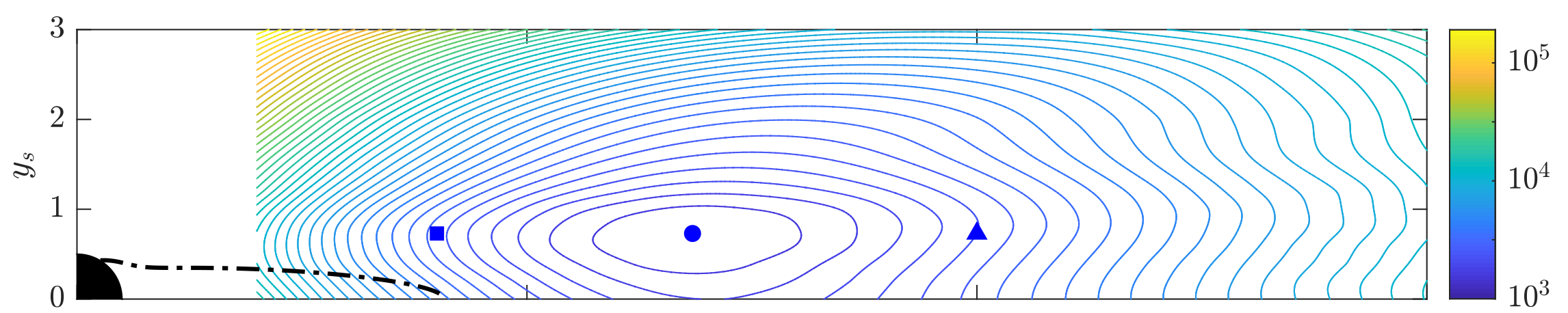}
    \llap{\parbox[b]{5.3in}{(a)\\\rule{0ex}{0.925in}}}
    }
    \centerline{
    \hspace{2mm}
    \includegraphics[width=0.975\textwidth]{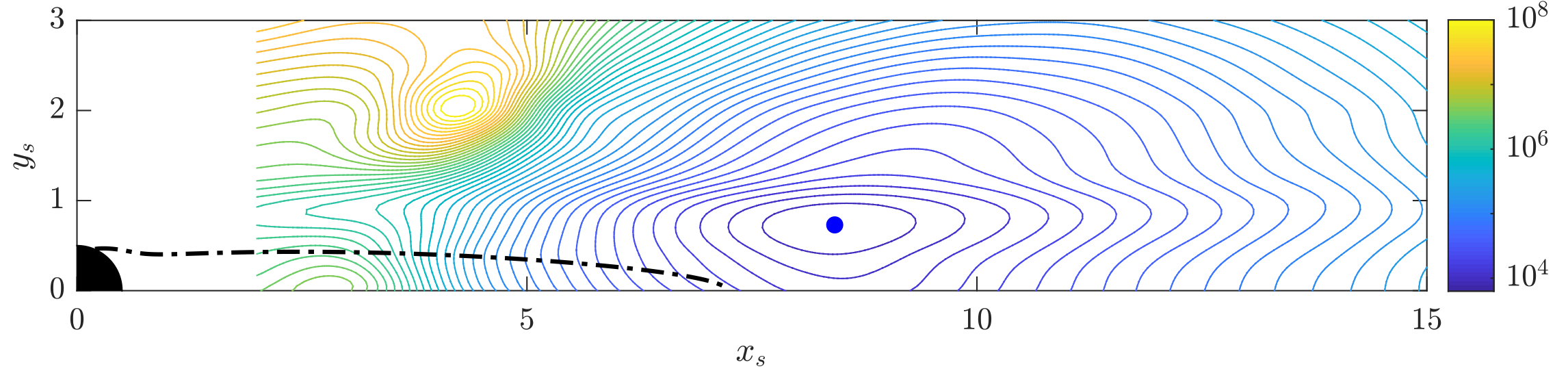}
    \llap{\parbox[b]{5.3in}{(b)\\\rule{0ex}{1.125in}}}
    }
    \caption[Brute-force sampling of sensor placements in the OE problem]{Contours of the cost function $\gamma^2{(k,\omega_n)}$ in the OE problem and the optimal sensor locations ($\protect\bluedot$) at (a) $\Rey=60$ and (b) $\Rey=110$. The square ($\protect\mysquare{blue}$) and triangle ($\protect\mytriangle{blue}$) correspond to the sensor locations considered in figure \ref{fig:lqe_sensorcomp_errdist}. Estimators are designed\protect\\using the parameters $k=3$ and $\omega_n=9$. The dash-dotted line (\protect\blacklineshort\hspace{0.5mm}\protect\blacksmalldot\hspace{0.5mm}\protect\blacklineshort) indicates the boundary of the reverse-flow region. }
    \label{fig:lqe_coarse_cont}
\end{figure}
To fully understand the effect of sensor placement, we start by performing a brute-force sampling approach for the OE problem at two Reynolds numbers: $\Rey=60$ and $\Rey=110$. The OE problem is solved by implementing the method introduced in \S\ref{sec:method} with parameters $k=3,\ \omega_n=9$. The corresponding cost function \eqref{equ:h2norm_resovent} is mapped out as a function of the sensor location ($x_{s},\ y_{s}$) in figure \ref{fig:lqe_coarse_cont}. As discussed in \S\ref{sec:design_method}, the cost function $\gamma_{(k,\omega_n)}^2$, though excluding the contribution from `background' modes, is sufficient to characterise the optimal estimator performance when random disturbances are applied everywhere, and thus to determine the global optimal sensor location.

Some critical features are immediately seen in figure \ref{fig:lqe_coarse_cont}. First of all, the global optimal sensor location ($\protect\bluedot$) is at approximately ($x_s$, $y_s$)=($6.83,\ 0.73$) for $\Rey=60$ and at ($x_s$, $y_s$)=($8.41,\ 0.73$) for $\Rey=110$. The optimal sensor location therefore moves downstream with increasing Reynolds number but its transverse position remains constant. Second, at the lower Reynolds number ($\Rey=60$), only one minimum exists. Although multiple extrema (maxima and minima) arise at the higher Reynolds number ($\Rey=110$), there is still only one local minimum in the wake area (downstream of the reverse-flow region) which is also the global minimum and far superior to any other minimum. Third, in both cases, placing the sensor too far upstream suffers a slightly higher penalty than placing it too far downstream. This is more clearly seen by plotting the spatial distribution of the estimation error throughout the domain. Thus, we define a root-mean-square value $\epsilon_{\textrm{OE}}$ such that:
\begin{equation}\label{equ:lqe_epsilon}
    \gamma^2{(k,\omega_n)}=\int_{\Omega}\epsilon_{\textrm{OE}}^2(x,y)\ d\Omega\ ,
\end{equation}
where $\epsilon_{\textrm{OE}}^2$ denotes the mean kinetic energy of the estimation error throughout the domain (see Appendix \ref{sec:app.c}). Figure \ref{fig:lqe_sensorcomp_errdist} uses the three sensor locations marked in figure \ref{fig:lqe_coarse_cont}(a), including the optimal location ($\protect\bluedot$), one located upstream ($\protect\mysquare{blue}$) and one located downstream ($\protect\mytriangle{blue}$), to show the effect of the sensor position on $\epsilon_{\textrm{OE}}$. 
\begin{figure}
    \centerline{
    \hspace{-110.75mm}
    \includegraphics[width=0.925\textwidth]{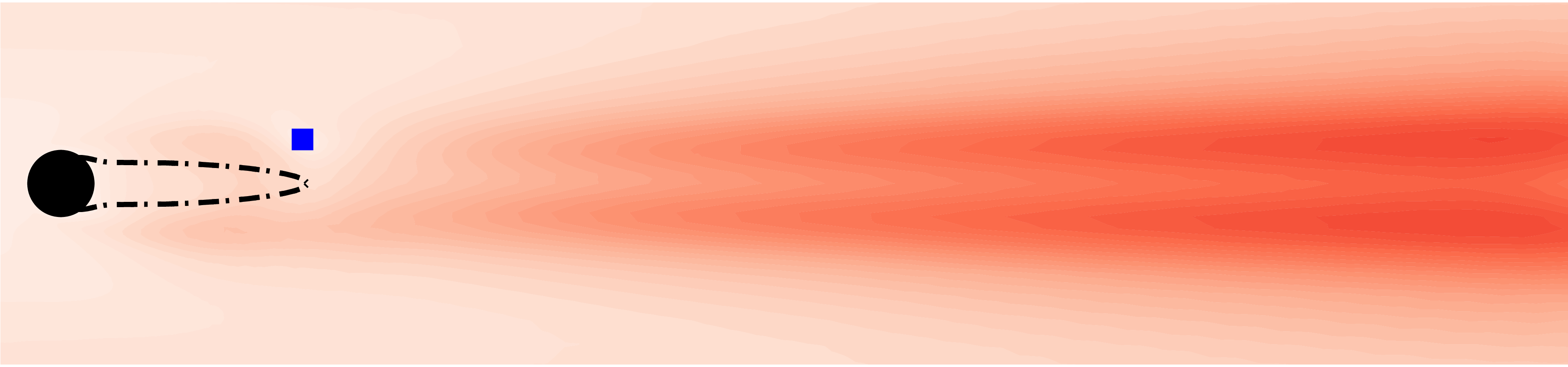}
    \llap{\parbox[b]{5.25in}{(a)\\\rule{0ex}{1.08in}}}  
    \hspace{-133.9mm}
    \includegraphics[trim=0 3 0 1,clip,width=0.1315\textwidth]{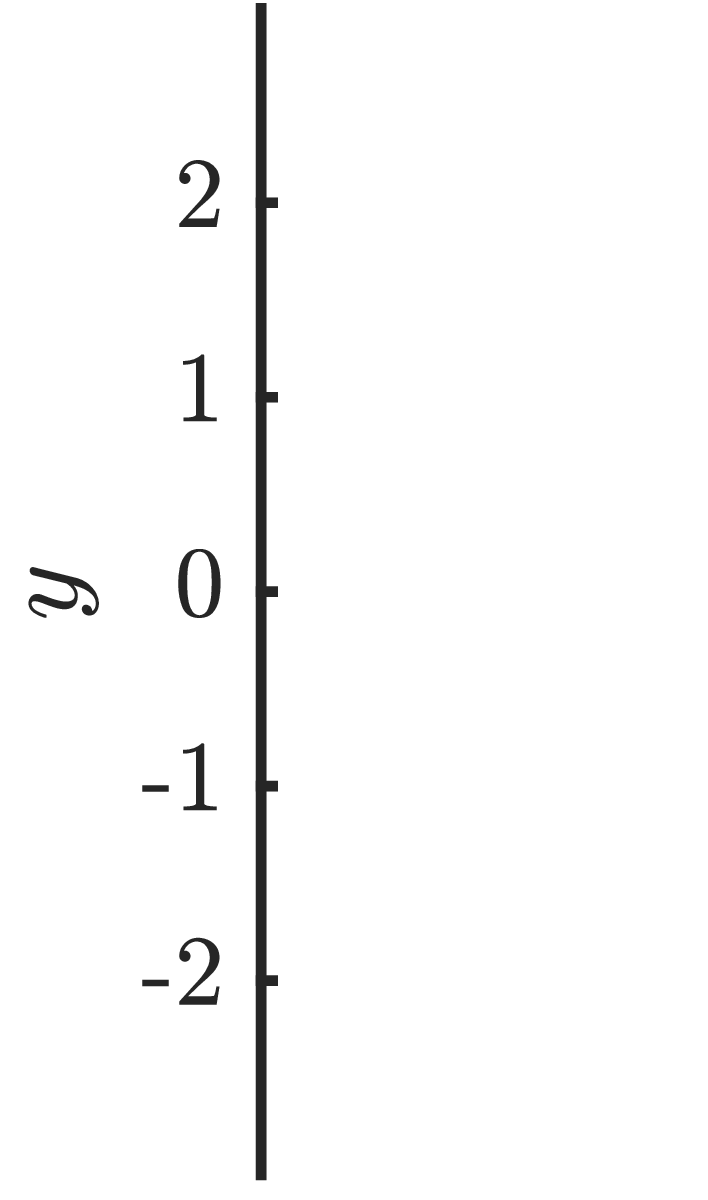}
    }
    \centerline{
    \hspace{-110.75mm}
    \includegraphics[width=0.925\textwidth]{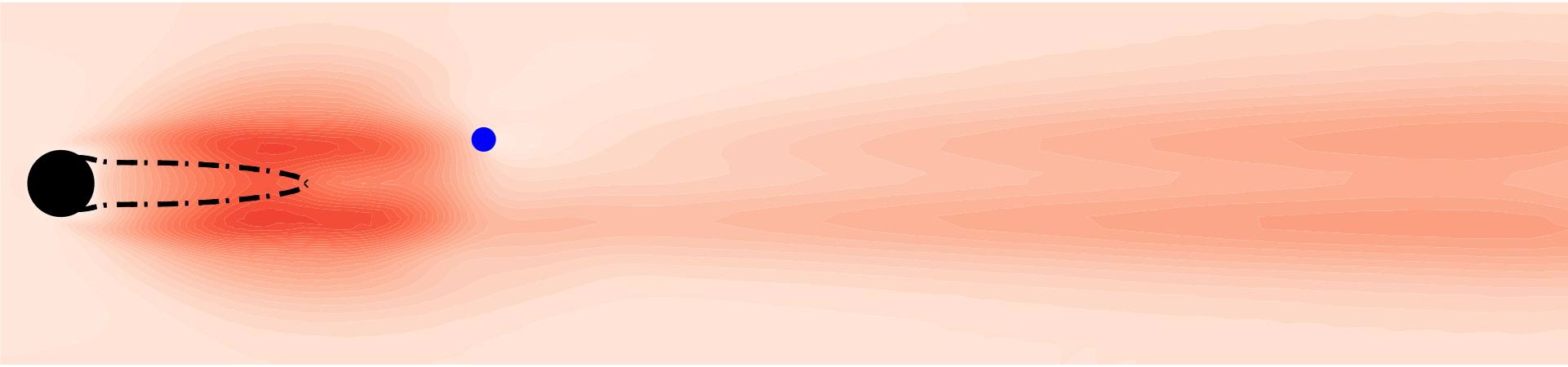}
    \llap{\parbox[b]{5.25in}{(b)\\\rule{0ex}{1.08in}}}  
    \hspace{-133.9mm}
    \includegraphics[trim=0 3 0 1,clip,width=0.1315\textwidth]{axis_y.png}
    }
    \centerline{
    \hspace{-110.75mm}
    \includegraphics[width=0.925\textwidth]{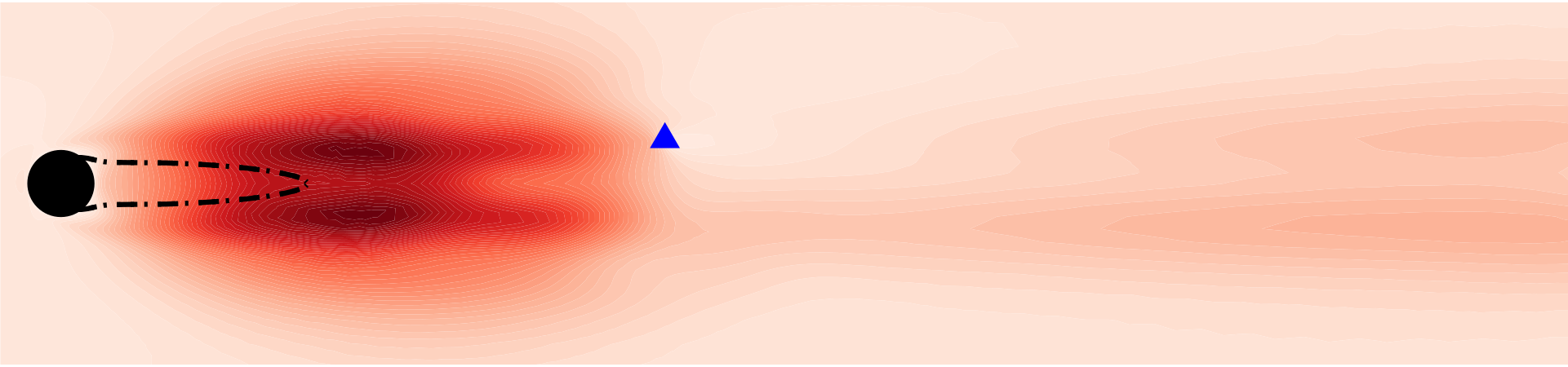}
    \llap{\parbox[b]{5.25in}{(c)\\\rule{0ex}{1.08in}}}  
    \hspace{-133.9mm}
    \includegraphics[trim=0 3 0 1,clip,width=0.1315\textwidth]{axis_y.png}
    }
    \vspace{-2.5mm}
    \centerline{
    \hspace{4.5mm}
    \includegraphics[width=0.94\textwidth]{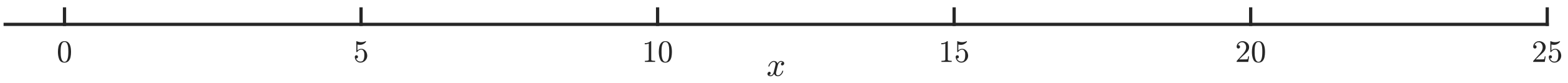}
    }
    \caption[Spatial distribution of the estimation error at $Re=60$]{Spatial distribution of the estimation error $\epsilon_{\textrm{OE}}$ at $\Rey=60$. Three different sensor locations shown in figure \ref{fig:lqe_coarse_cont}(a) are tested, which are marked by (a) $\protect\mysquare{blue}$; (b) $\protect\bluedot$ (optimal); (c) $\protect\mytriangle{blue}$. The optimal estimators are designed using the parameters $k=3$ and $\omega_n=9$. The dash-dotted line (\protect\blacklineshort\hspace{0.5mm}\protect\blacksmalldot\hspace{0.5mm}\protect\blacklineshort) indicates the boundary of the reverse-flow region and all \protect\\plots share the same linear colour scale.}
    \label{fig:lqe_sensorcomp_errdist}
\end{figure}

In all three cases, the smallest value of $\epsilon_{\textrm{OE}}$ occurs at the sensor location. The most significant contributions to the estimation error are concentrated in two horizontal streaks which are approximately symmetric and located around $y\approx 0.7$. The reduction of $\epsilon_{\textrm{OE}}$ at the sensor location divides these two streaks into two regions: a near-wake area (between the cylinder and the sensor) and a far-wake area (downstream of the sensor). If the sensor is placed too far upstream, the upstream estimation error is naturally dampened but the downstream flow is not observable to the sensor and thus the estimation error develops in the large far-wake area. When the sensor moves downstream, the estimation error is strongly amplified in the near-wake area, as shown in figure \ref{fig:lqe_sensorcomp_errdist}(c). The optimal placement of the sensor should therefore balance minimising the estimation error that is amplified in the near-wake area against minimising that developing in the far-wake area, which agrees with the findings for a spatially developing one-dimensional flow  \citep{oehler2018sensor}. 
\subsubsection{Optimal sensor placement}
\begin{figure}
    \vspace{1mm}
    \centerline{
    \includegraphics[width=0.48\textwidth]{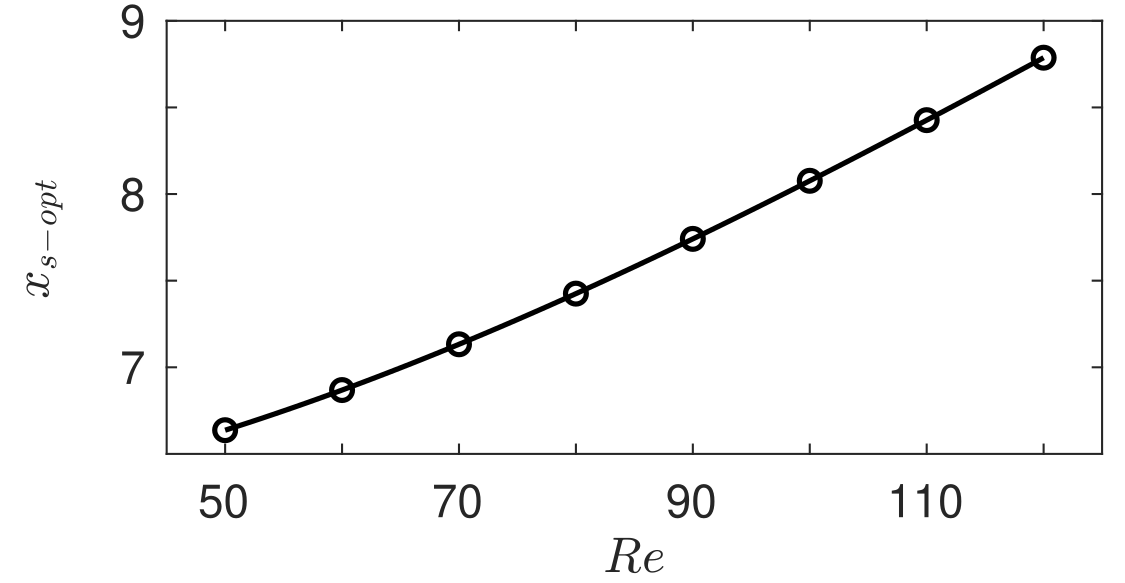}
    \llap{\parbox[b]{2.65in}{(a)\\\rule{0ex}{1.225in}}}
    \includegraphics[width=0.48\textwidth]{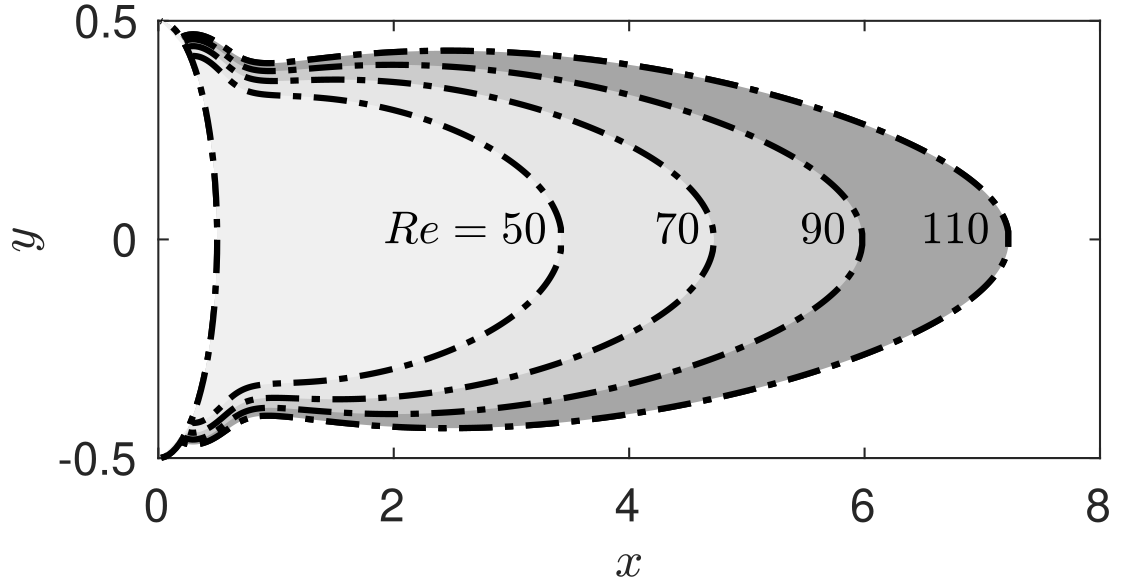}
    \llap{\parbox[b]{2.65in}{(c)\\\rule{0ex}{1.225in}}}
    }
    \centerline{
    \includegraphics[width=0.48\textwidth]{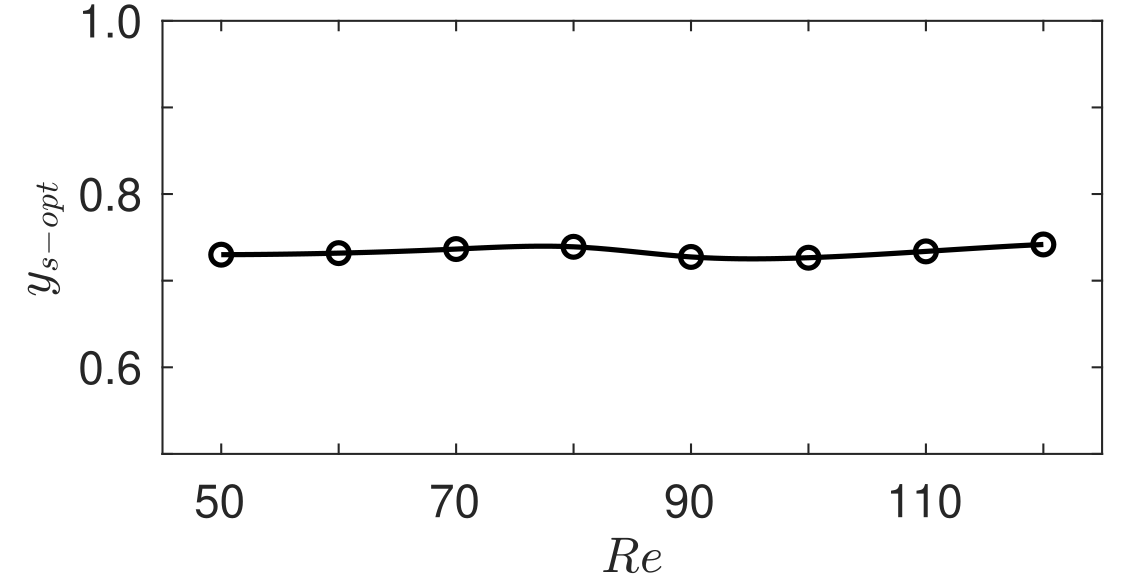}
    \llap{\parbox[b]{2.65in}{(b)\\\rule{0ex}{1.225in}}}
    \includegraphics[width=0.48\textwidth]{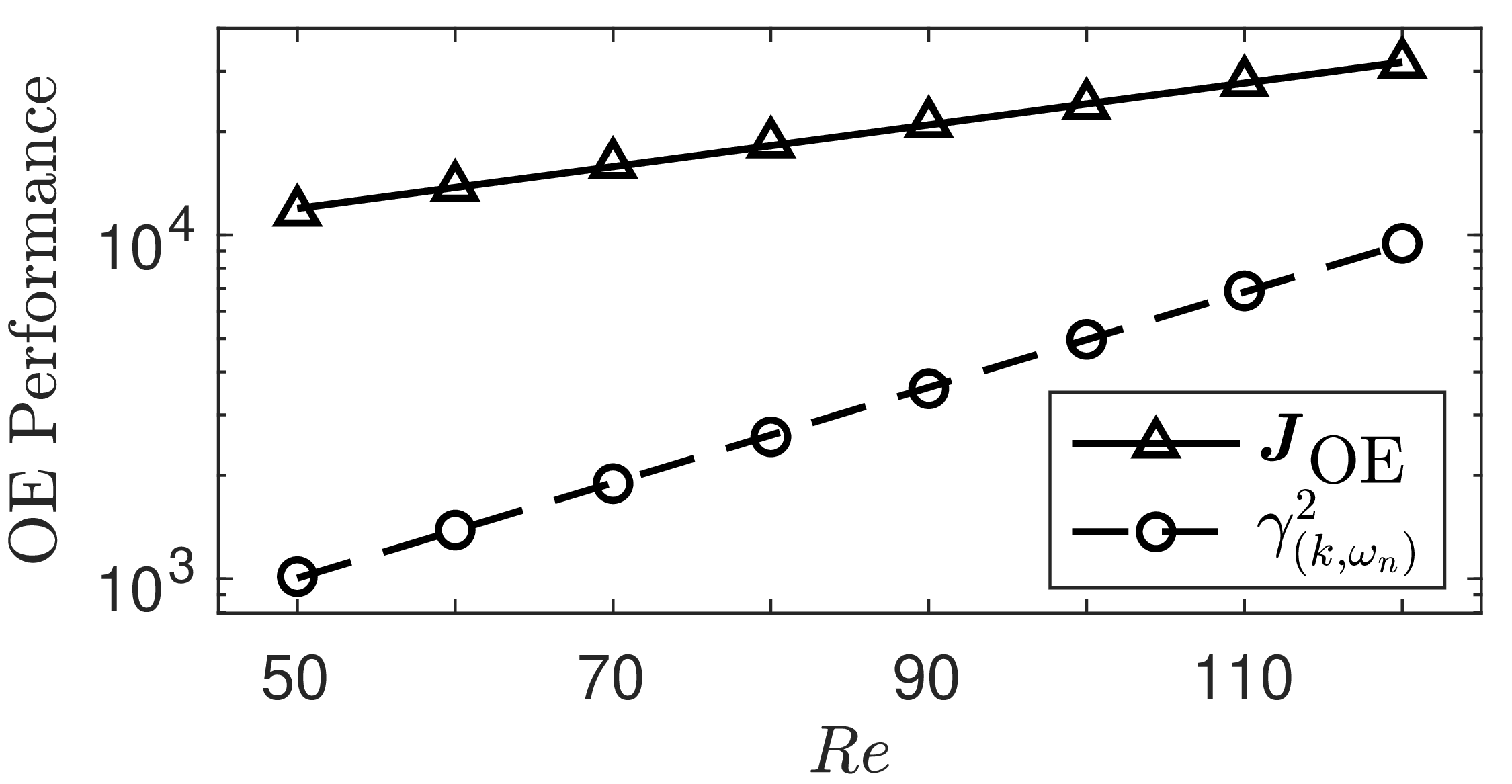}
    \llap{\parbox[b]{2.65in}{(d)\\\rule{0ex}{1.225in}}}
    }
    \caption[The optimal sensor placement results for the OE problem]{Coordinates of the optimal sensor location (a) $x_{s-opt}$ and (b) $y_{s-opt}$ as a function of the Reynolds number. (c) The profile of the reverse-flow region (the grey area) at different Reynolds numbers. (d) The optimal estimation performance as a function of the \protect\\Reynolds number. $\gamma^2{(k,\omega_n)}$ is the cost function for the estimator design ($k=3$, $\omega_n=9$) whereas $\textbf{\textit{J}}_\textrm{OE}$ represents the mean kinetic energy of the total estimation error evaluated from numerical simulations with random disturbances applied everywhere.}
    \label{fig:lqe_optlocsh2n}
\end{figure}
We are interested in the optimal sensor locations to achieve the best estimation performance for different Reynolds numbers. Brute-force sampling is inefficient when searching for global optimal placements for different Reynolds numbers. Instead, we now employ a gradient minimisation method (e.g.~Newton's method) together with sensible initial guesses chosen based on the observations of figure \ref{fig:lqe_coarse_cont}. 

Coordinates of the optimal sensor location are plotted in figure \ref{fig:lqe_optlocsh2n}(a, b) as a function of the Reynolds number. It can be seen that, as Reynolds number increases, the optimal streamwise location $x_{s-opt}$ moves downstream whereas the optimal transverse location $y_{s-opt}$ remains approximately constant ($0.73\pm0.01$). This trend is consistent with the evolution of the reverse-flow region shown in figure \ref{fig:lqe_optlocsh2n}(c), which also closely approximates the absolutely unstable (AU) region \citep{pier2002frequency}. The length of the AU region behind the circular cylinder increases with increasing Reynolds number, which pushes the optimal sensor location downstream. However, to avoid excessive time lag, the migration of the optimal sensor location downstream is much slower than the evolution of the AU region: the optimal sensor location moves only $1.8$ diameters downstream whereas the length of the AU region extends $3.8$ diameters further when the Reynolds number increases from $50$ to $110$. This is a result of the convection-driven nature of the system: at higher Reynolds numbers, the instability develops more rapidly while convecting downstream and thus there exists a larger effective phase lag in the measurements when the sensor is placed at a fixed distance downstream of the AU region. This is consistent with the findings of \cite{oehler2018sensor}, where the optimal sensor locations found for the OE problem moved upstream to compensate for any artificial time lag that was imposed on the system.

It is also interesting to note that, although the AU region extends further downstream with increasing Reynolds number, it barely changes in the transverse direction. As a result, the optimal transverse position remains almost constant over the Reynolds number range considered. Meanwhile, the cost function $\gamma^2{(k,\omega_n)}$ for the estimator design, which considers only the first $k$ resolvent modes over the frequency range $\omega\in [-\omega_n, \omega_n]$, increases logarithmically, as shown in figure \ref{fig:lqe_optlocsh2n}(d). To evaluate the optimal estimation performance when random disturbances are applied everywhere, we simulate the closed-loop error system with the optimal sensor placement and the optimal estimator. The mean kinetic energy of the total estimation error $\textit{\textbf{J}}_\textrm{OE}$, as defined by equation \eqref{equ:cost_j}, is plotted as a function of the Reynolds number in figure \ref{fig:lqe_optlocsh2n}(d). Similarly, the estimation performance $\textit{\textbf{J}}_\textrm{OE}$ also increases logarithmically with increasing Reynolds number. The gap between $\gamma^2{(k,\omega_n)}$ and $\textit{\textbf{J}}_\textrm{OE}$ represents the contribution from the `background' or `freestream' modes that are excluded while designing the optimal estimator, which accounts for $91\%$ of the total estimation error at $\Rey=50$ but decreases to $75\%$ at $\Rey=110$. Note that the solution of the optimisation problem (e.g.~optimal estimator, optimal sensor placement) is determined by the gradient of the cost function (i.e.~by setting the gradient equal to zero). Therefore, although there is a large gap between $\gamma^2{(k,\omega_n)}$ and $\textit{\textbf{J}}_\textrm{OE}$, both of them give the same optimal location.

\subsubsection{Effect of domain size}\label{sec:lqe_domain_size}
\begin{figure}
    \centerline{
    \includegraphics[width=0.48\textwidth]{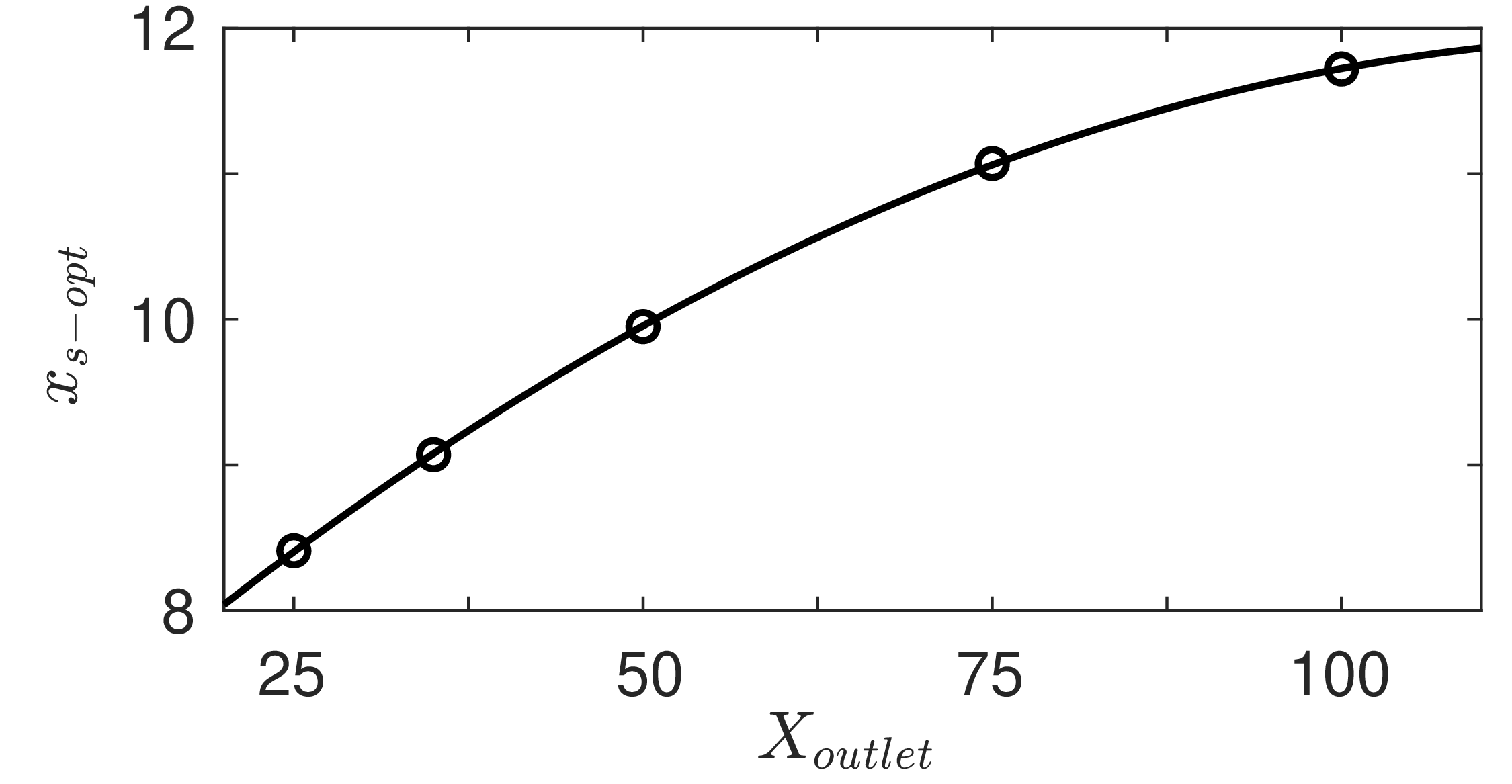}
    \llap{\parbox[b]{2.65in}{(a)\\\rule{0ex}{1.2in}}}
    \includegraphics[width=0.48\textwidth]{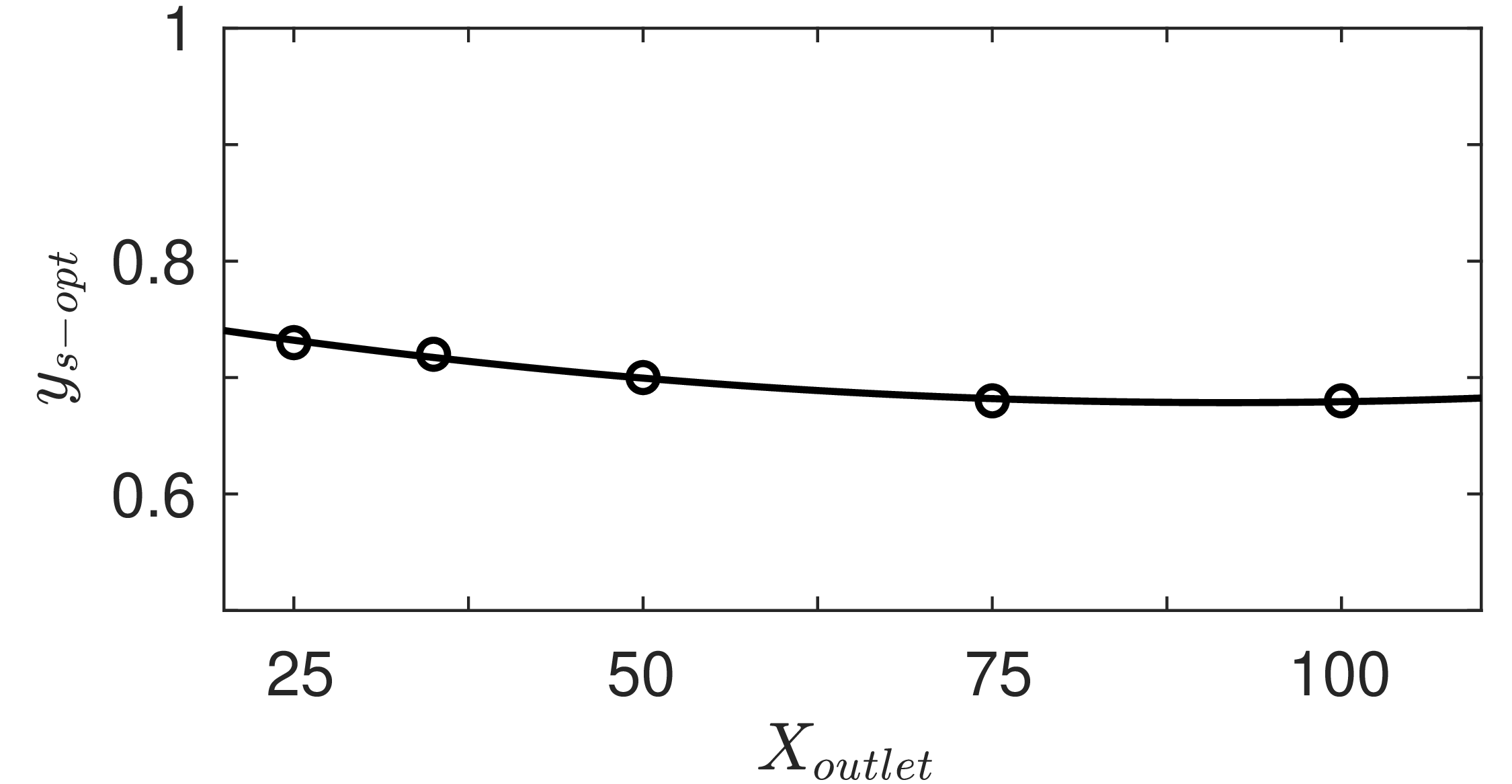}
    \llap{\parbox[b]{2.65in}{(b)\\\rule{0ex}{1.2in}}}
    }
    \caption[The optimal sensor location in the OE problem using extended domains.]{Coordinates of the optimal sensor location (a) $x_{s-opt}$ and (b) $y_{s-opt}$ at $\Rey$=110 when the domain is extended in the streamwise direction. $X_{outlet}$ is the position of the outlet boundary.}
    \label{fig:lqe_optsensor_mesh}
\end{figure}
In the OE problem, we simply choose to optimally estimate the entire flow field, which is influenced by two different mechanisms: i) the growth of flow perturbations due to the strong shear layer in the near-wake area of the base flow; ii) the transportation of flow perturbations to the far-wake area due to the convective nature of the flow. Due to the lack of nonlinear energy transfer to higher frequencies, any flow perturbations will grow and be transported far downstream of the cylinder until they dissipate. However, this leads to two significant problems. First, in contrast to the optimal placements in other control problems, the optimal sensor location in the OE problem is sensitive to the streamwise extent of the computational domain. When the domain is extended in the streamwise direction, i.e.~estimation extends further downstream, the optimal sensor location also moves downstream. This can be clearly seen in figure \ref{fig:lqe_optsensor_mesh} where the optimal sensor location at $\Rey=110$ is plotted as a function of the streamwise extent of the domain. In particular, the streamwise coordinate $x_{s-opt}$ does not converge even when the domain extends to 100 diameters downstream of the cylinder. Second, the optimal sensor placement found for the OE problem is not optimal for feedback control. Intuitively, the upstream growth of flow perturbations should be the main concern for feedback control and the control of their convection downstream would not be effective at minimising flow perturbations. Indeed, we will see later that the optimal locations of either the actuator or the sensor found for feedback control are always upstream or near the edge of the reverse-flow region whereas the sensor in the OE problem is best-placed far downstream. Together these observations indicate that the OE problem, although interesting, is not the most suitable method for placing sensors for feedback control. Therefore, effective estimation of the flow field does not guarantee effective control performance and vice versa. Similar conclusions have been drawn while investigating the sensor placement problem for a one-dimensional flow \citep{oehler2018sensor} and for the cylinder flow using deep reinforcement learning \citep{paris2021robust}. Note that the OE problem considered in this study is merely a part of the whole optimal feedback control problem. Although the optimal sensor placement found for the OE problem varies with the domain size, the computational domain described in figure \ref{fig:compudomain} results in converged optimal placements for the other control problems considered.

\subsection{Full-state information control problem}
\subsubsection{Brute-force sampling}\label{sec:lqr_brute}
\begin{figure}
    \vspace{2mm}
    \centerline{
    \hspace{2mm}
    \includegraphics[width=0.47\textwidth]{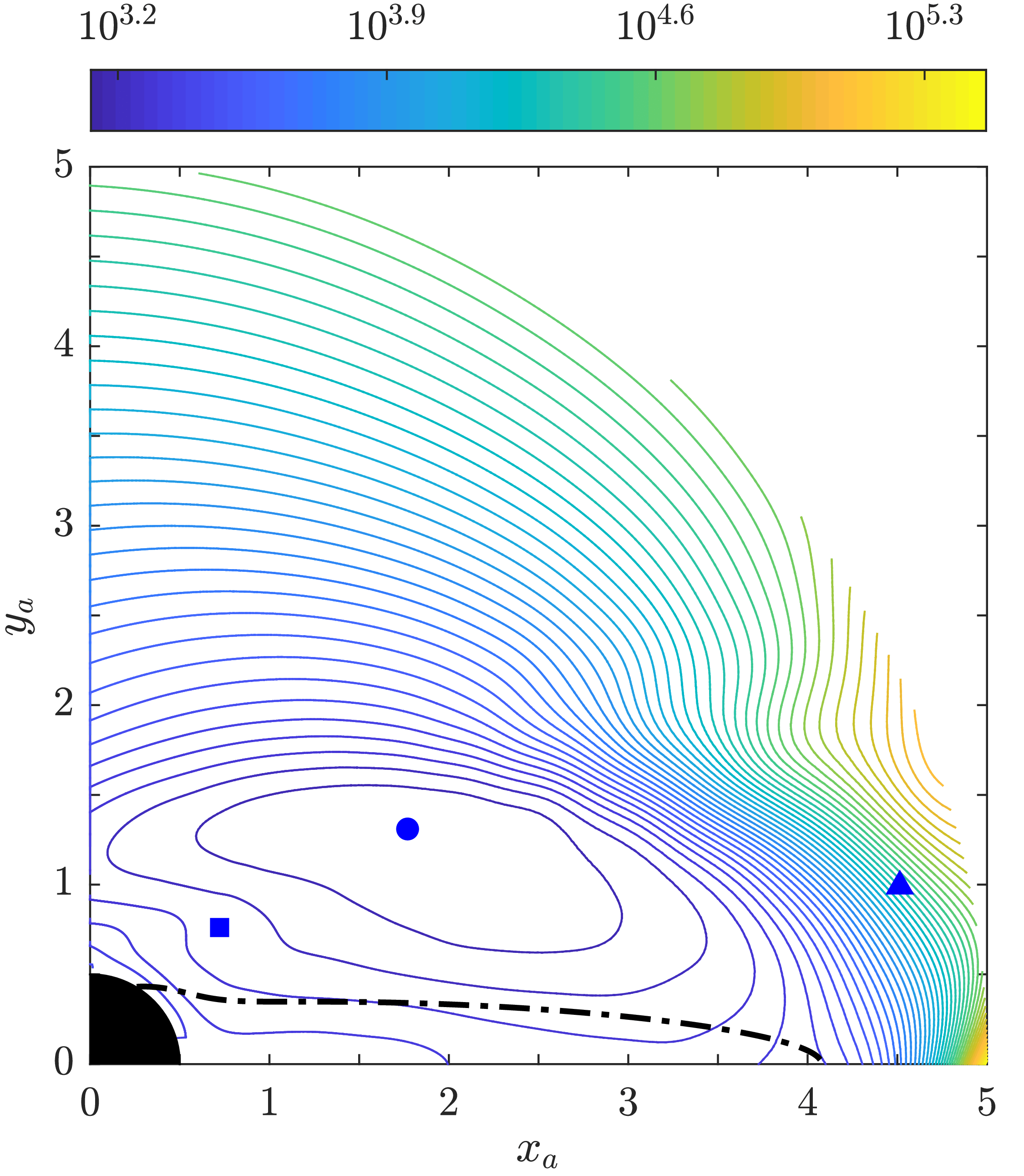}
    \llap{\parbox[b]{2.7in}{(a)\\\rule{0ex}{2.75in}}}
    \includegraphics[width=0.47\textwidth]{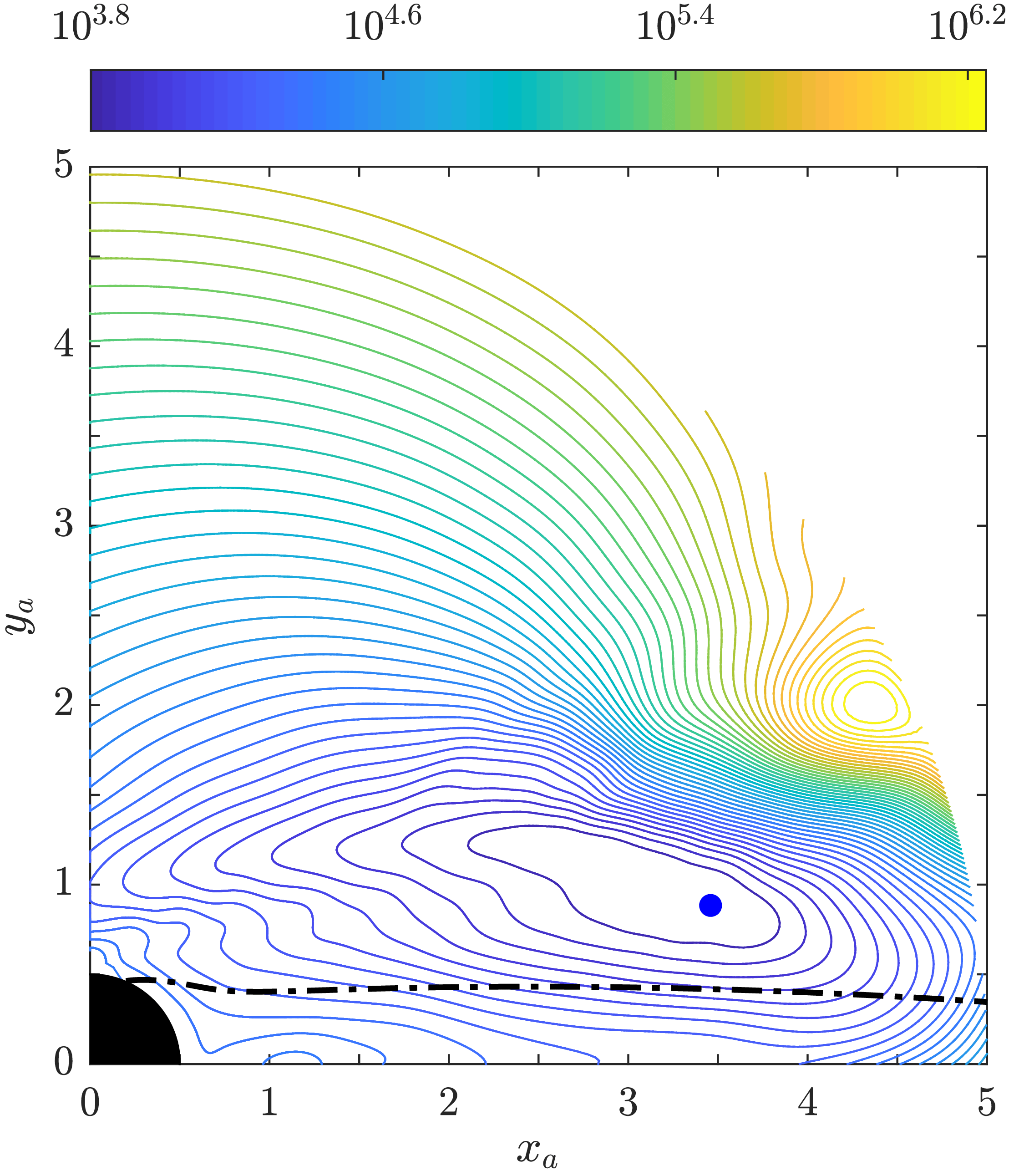}
    \llap{\parbox[b]{2.7in}{(b)\\\rule{0ex}{2.75in}}}
    }
    \caption[Brute-force sampling of actuator placements in the FIC problem]{Contours of the cost function $\gamma^2{(k,\omega_n)}$ in the FIC problem and the optimal actuator locations ($\protect\bluedot$) at (a) $\Rey=60$ and (b) $\Rey=110$. The square ($\protect\mysquare{blue}$) and triangle ($\protect\mytriangle{blue}$) correspond to the actuator locations considered in figure \ref{fig:lqr_actuatorcomp_xdist}. Controllers are designed using the parameters $k=3$ and $\omega_n=9$. The dash-dotted line (\protect\blacklineshort\hspace{0.5mm}\protect\blacksmalldot\hspace{0.5mm}\protect\blacklineshort) indicates the boundary of the reverse-flow region.}
    \label{fig:lqr_coarse_cont}
\end{figure}

We now turn our attention to the FIC problem and the corresponding optimal actuator placement problem. Analogous to the OE problem, a brute-force sampling approach for the FIC problem is performed at $\Rey=60$ and $\Rey=110$ with the same parameters ($k=3$, $\omega_n=9$) as those used in the OE problem. The corresponding cost function $\gamma^2{(k,\omega_n)}$ is mapped as a function of the actuator location ($x_{a},\ y_{a}$) in figure \ref{fig:lqr_coarse_cont}, where the dash-dotted line indicates the region of reverse flow.

\begin{figure}
    \centerline{
    \hspace{-110.75mm}
    \includegraphics[width=0.925\textwidth]{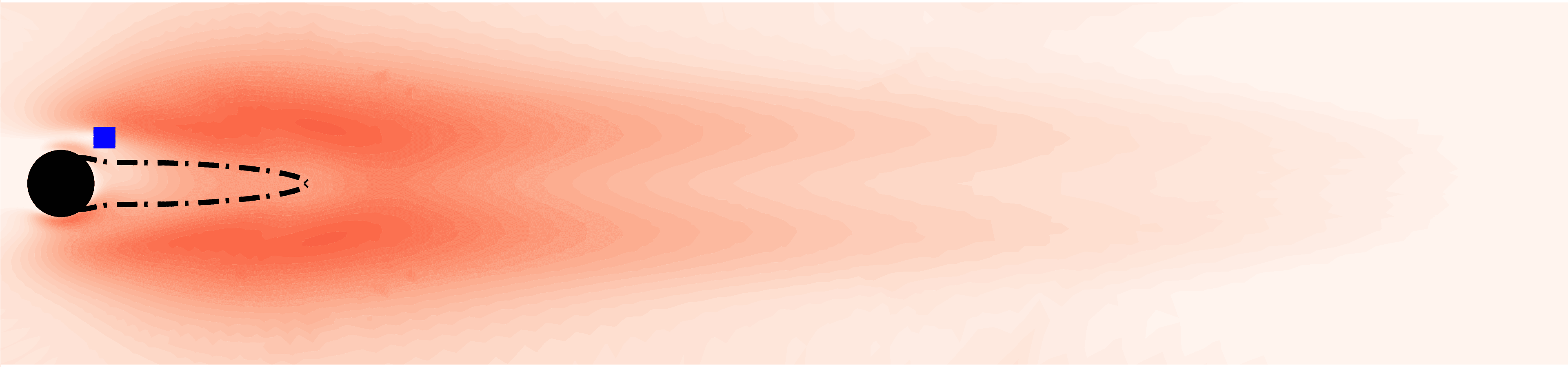}
    \llap{\parbox[b]{5.25in}{(a)\\\rule{0ex}{1.08in}}}  
    \hspace{-133.9mm}
    \includegraphics[trim=0 3 0 1,clip,width=0.1315\textwidth]{axis_y.png}
    }
    \centerline{
    \hspace{-110.75mm}
    \includegraphics[width=0.925\textwidth]{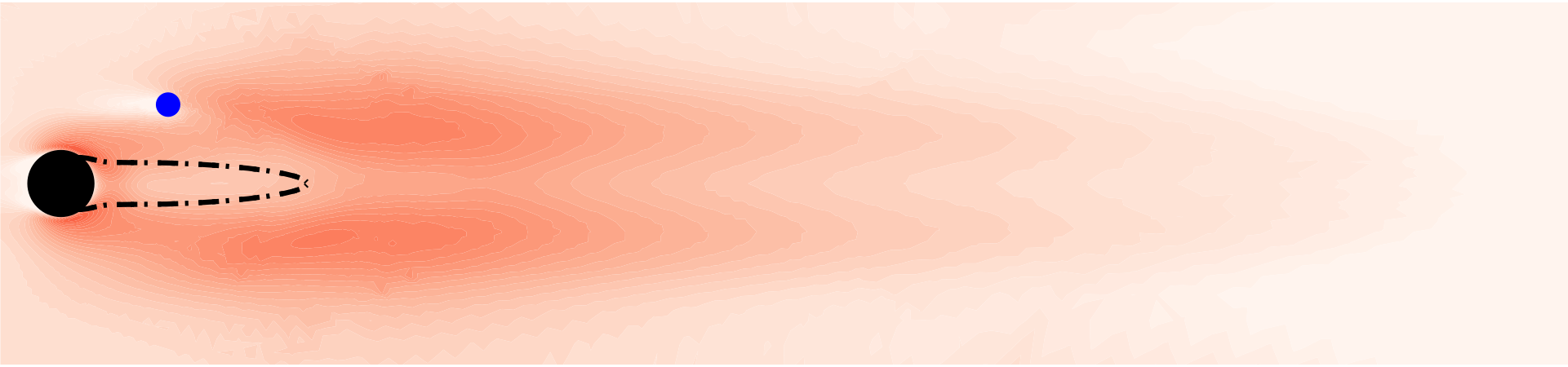}
    \llap{\parbox[b]{5.25in}{(b)\\\rule{0ex}{1.08in}}}  
    \hspace{-133.9mm}
    \includegraphics[trim=0 3 0 1,clip,width=0.1315\textwidth]{axis_y.png}
    }
    \centerline{
    \hspace{-110.75mm}
    \includegraphics[width=0.925\textwidth]{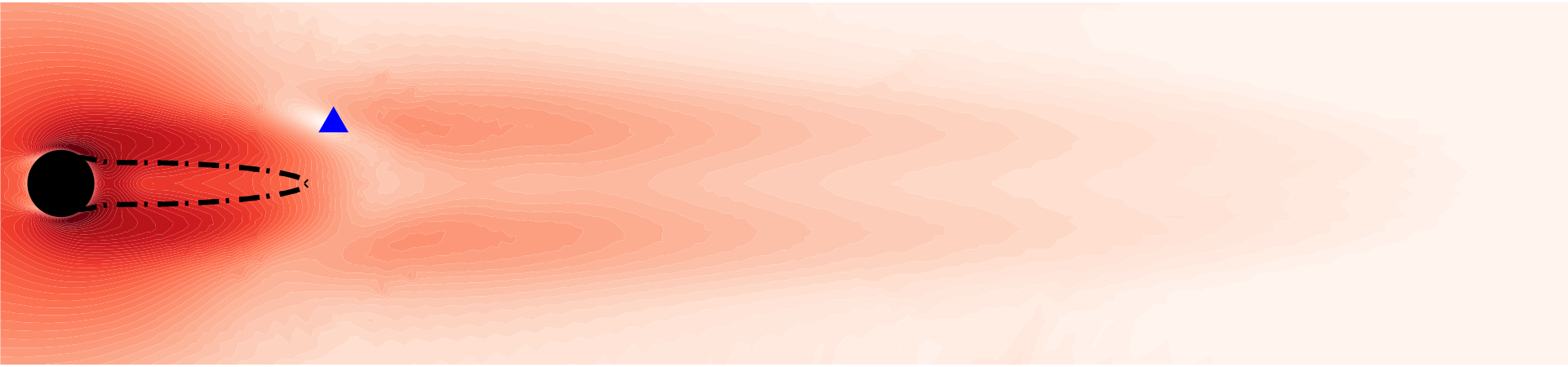}
    \llap{\parbox[b]{5.25in}{(c)\\\rule{0ex}{1.08in}}}  
    \hspace{-133.9mm}
    \includegraphics[trim=0 3 0 1,clip,width=0.1315\textwidth]{axis_y.png}
    }
    \vspace{-2.5mm}
    \centerline{
    \hspace{4.5mm}
    \includegraphics[width=0.94\textwidth]{axis_x.png}
    }
    \caption[Spatial distribution of the receptivity to disturbance at $Re=60$]{Spatial distribution of the receptivity to disturbance $\epsilon_{\textrm{FIC}}$ at $\Rey=60$. Three different actuator locations shown in figure \ref{fig:lqr_coarse_cont}(a) are tested, which are marked by (a) $\protect\mysquare{blue}$;\protect\\(b) $\protect\bluedot$ (optimal); (c) $\protect\mytriangle{blue}$. The optimal controllers are designed with $k=3$ and $\omega_n=9$. The dash-dotted line (\protect\blacklineshort\hspace{0.5mm}\protect\blacksmalldot\hspace{0.5mm}\protect\blacklineshort) indicates the boundary of the reverse-flow region and all plots share the same logarithmic colour scale.}
    \label{fig:lqr_actuatorcomp_xdist}
\end{figure}

We first notice that the global optimal actuator ($\protect\bluedot$) is located at approximately ($x_a$, $y_a$)=($1.77,\ 1.31$) for $\Rey=60$ and at ($x_a$, $y_a$)=($3.46,\ 0.88$) for $\Rey=110$. Similar to the OE problem, only one minimum appears in the sampled area at each Reynolds number, which is therefore the global optimum. The performance of an FIC controller with the actuator placed outside the sampled area is far worse than that with an actuator placed inside the sampled area. It is reasonable to suppose that the optimal actuator locations for the Reynolds numbers considered in this study will always be in the near-wake area and outside the reverse-flow region. As can be seen from figure \ref{fig:lqr_coarse_cont}, we expect that, with increasing Reynolds number, the optimal actuator placement will move not only downstream but also closer to the reverse-flow region. These observations provide reasonable initial guesses for the optimal placement problem at different Reynolds numbers and a traditional gradient minimisation method (e.g.~Newton’s method) is sufficient to solve this non-convex problem. Another significant feature observed from the topography of the cost function in figure \ref{fig:lqr_coarse_cont} is that a `cliff' appears slightly downstream of the optimal actuator position at each Reynolds number, where the cost function increases rapidly. This occurs because the shift in the actuator location across the `cliff' causes a right-half-plane (RHP) zero to appear near the unstable pole in the $\textbf{\textit{q}}$-to-$\textit{\textbf{y}}$ transfer function. This imposes a severe limitation on the controller's ability to reject disturbances (see \cite{skogestad2007multivariable} and the discussion of figure \ref{fig:lqr_actuatorcomp_xdist}). 

In order to clearly show the controller's disturbance rejection performance for different actuator placements, we plot the root-mean-square value $\epsilon_{\textrm{FIC}}$ in figure \ref{fig:lqr_actuatorcomp_xdist}, which is defined similarly to equation \eqref{equ:lqe_epsilon} (see Appendix \ref{sec:app.c}). Analogous to the OE problem, we only consider the first $k$ resolvent modes within the frequency range $\omega\in [-\omega_n,\omega_n]$ to make the computation tractable. Here, the term $\epsilon_{\textrm{FIC}}^2$ indicates the contribution of each and every disturbance to the mean kinetic energy of flow perturbations under closed-loop control. In other words, it shows the spatial distribution of the receptivity of the closed-loop system to disturbances. The darker regions in figure \ref{fig:lqr_actuatorcomp_xdist} have higher receptivity, which implies that the flow is sensitive to disturbances (poorer disturbance-rejection ability) under the control of the actuator.

In order to show the effect of actuator position on $\epsilon_{\textrm{FIC}}$, figure \ref{fig:lqr_actuatorcomp_xdist} uses the three actuator locations marked in figure \ref{fig:lqr_coarse_cont}(a): the optimal location ($\protect\bluedot$), one located upstream ($\protect\mysquare{blue}$) and one located downstream ($\protect\mytriangle{blue}$). The same logarithmic colour scale is used for all three contour plots. In each case, the minimum value of $\epsilon_{\textrm{FIC}}$ occurs at the location of the actuator, which is where the disturbance's influence can be directly eliminated. We also see that, for all three plots, there is a small white area upstream of the actuator rather than downstream of it. Therefore, the actuator is better able to reject the influences of disturbances that occur immediately upstream of the actuator than those that occur immediately downstream. Another significant observation from these contour plots is that the spatial distribution of $\epsilon_{\textrm{FIC}}$ can also be divided into two regions: the first is the near-wake area (upstream of the actuator) and the second is the far-wake area (downstream of the actuator). If the actuator is placed too close to the cylinder, it successfully rejects upstream disturbances but fails to suppress disturbances in the large far-wake area, as shown by figure \ref{fig:lqr_actuatorcomp_xdist}(a). But if the actuator is moved to the far-wake area, the situation becomes much worse, as shown in figure \ref{fig:lqr_actuatorcomp_xdist}(c), where the influence of upstream disturbances is much greater than for the other two cases. This occurs because the actuator is located inside the `cliff' region presented in figure \ref{fig:lqr_coarse_cont}(a). The optimal actuator position should therefore allow the actuator to handle the effect of disturbances from both the near-wake area and the far-wake area, which is consistent with previous results for the one-dimensional Ginzburg-Landau equation \citep{oehler2018sensor}. 

\subsubsection{Optimal actuator placement}
\begin{figure}
    \vspace{2mm}
    \centerline{
    \includegraphics[width=0.48\textwidth,valign=b]{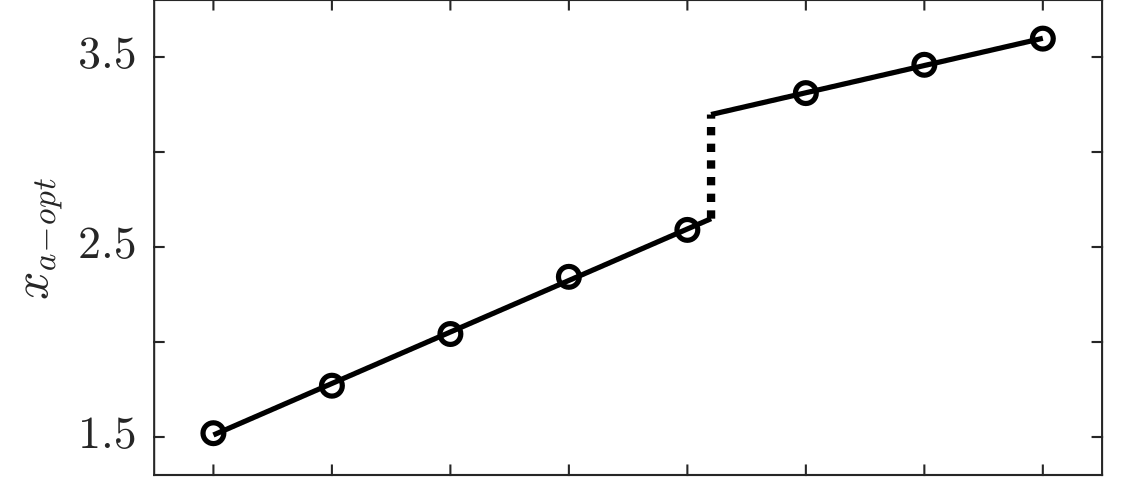}
    \llap{\parbox[b]{2.625in}{(a)\\\rule{0ex}{1in}}}
    \includegraphics[width=0.48\textwidth,valign=b]{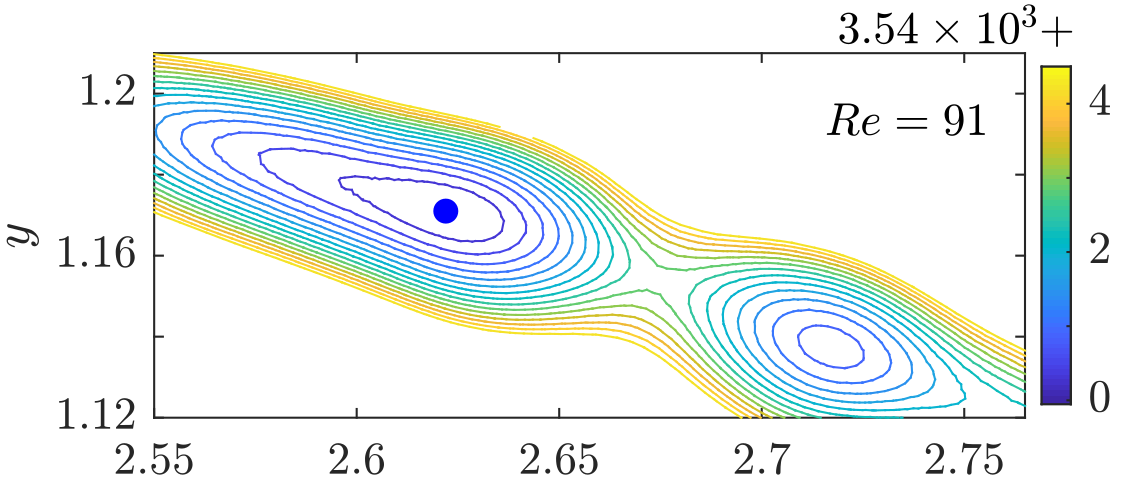}
    \llap{\parbox[b]{2.625in}{(d)\\\rule{0ex}{1in}}}
    }
    \centerline{
    \includegraphics[width=0.48\textwidth,valign=b]{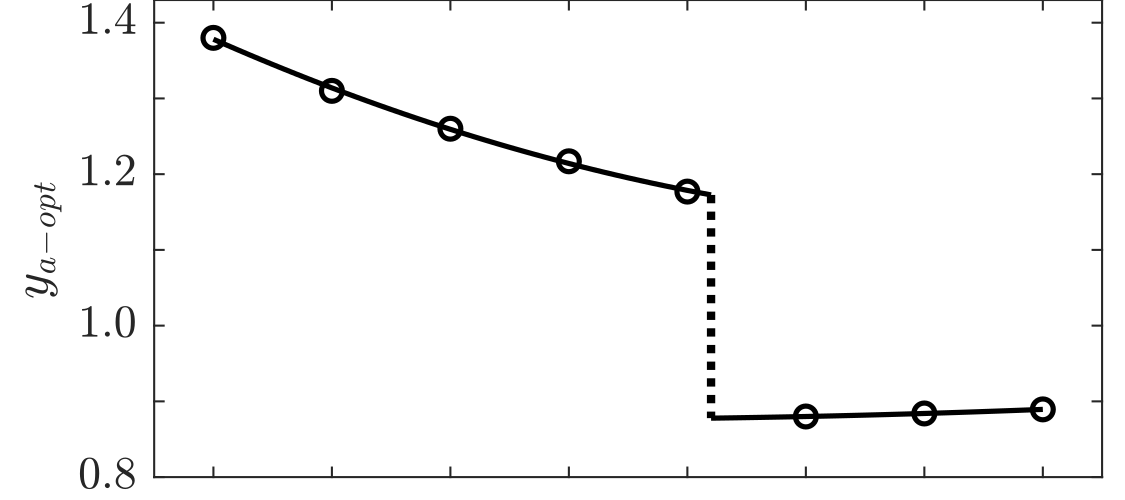}
    \llap{\parbox[b]{2.625in}{(b)\\\rule{0ex}{1in}}}
    \includegraphics[width=0.48\textwidth,valign=b]{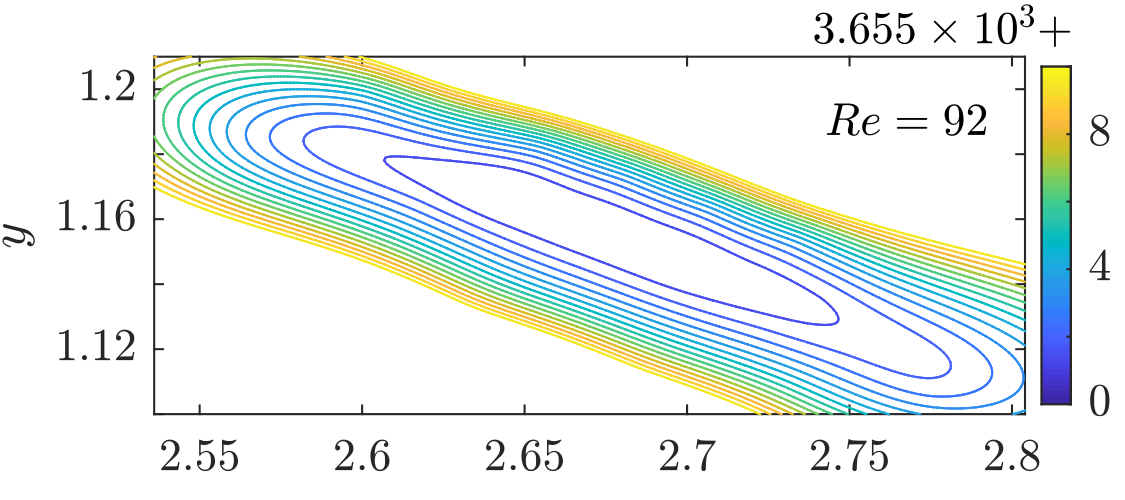}
    \llap{\parbox[b]{2.625in}{(e)\\\rule{0ex}{1in}}}
    }
    \centerline{
    \includegraphics[width=0.48\textwidth,valign=b]{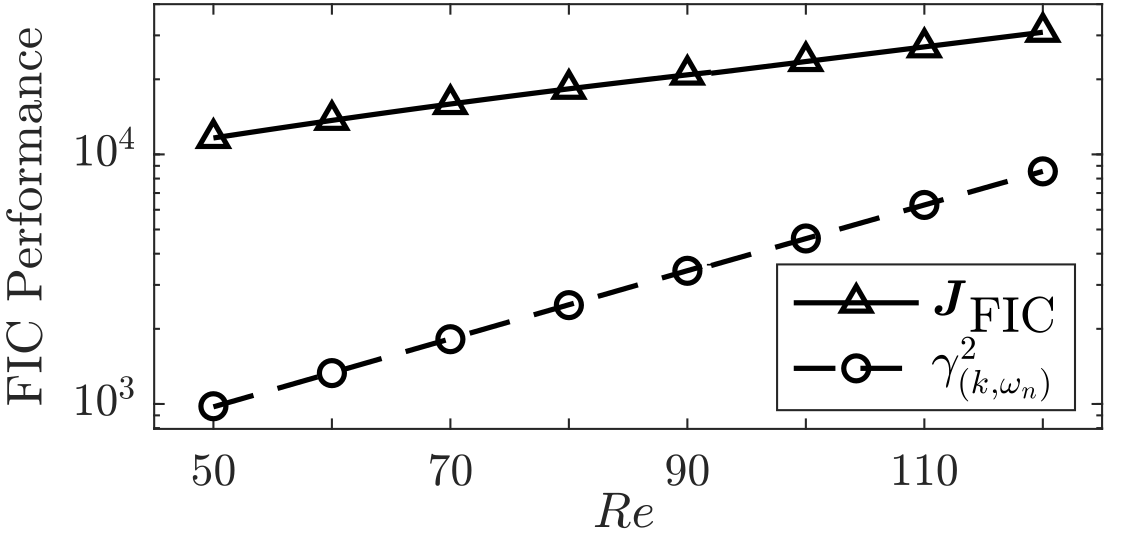}
    \llap{\parbox[b]{2.625in}{(c)\\\rule{0ex}{1.175in}}}
    \includegraphics[width=0.48\textwidth,valign=b]{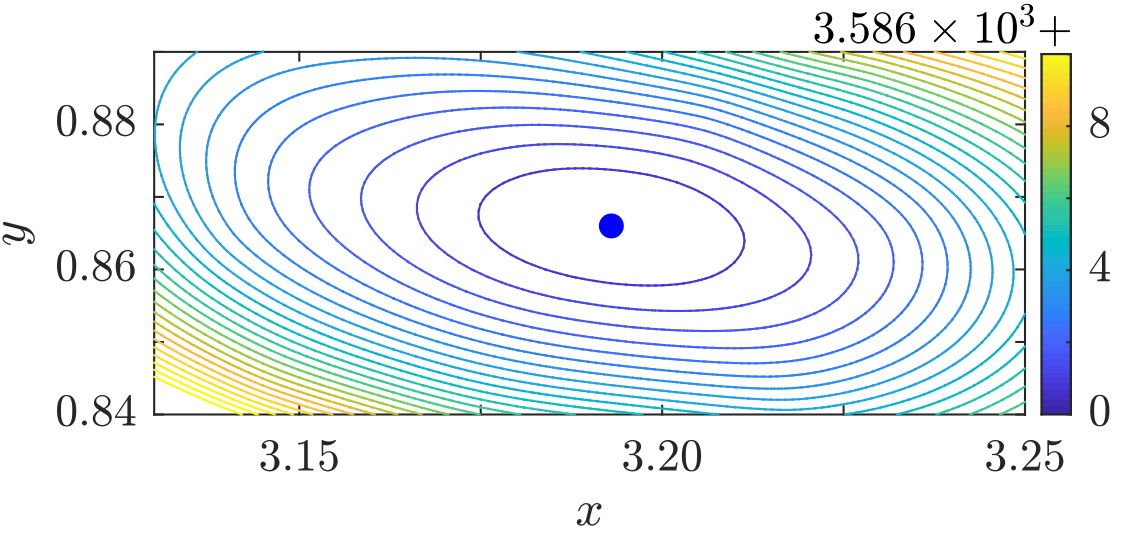}
    \llap{\parbox[b]{2.625in}{(f)\\\rule{0ex}{1.175in}}}
    }
    \caption[Optimal actuator placements and performances for the FIC problem]{Left: Coordinates of the optimal actuator location (a) $x_{a-opt}$ and (b) $y_{a-opt}$ as a function of the Reynolds number. (c) The optimal FIC control performance as a function of the Reynolds number. $\gamma^2{(k,\omega_n)}$ is the cost function for the controller design ($k=3$, $\omega_n=9$) whereas $\textbf{\textit{J}}_\textrm{FIC}$ is the mean kinetic energy of flow perturbations from numerical simulations where random disturbances are applied everywhere. A discontinuity in the optimal actuator location  ($\protect\dottedline$) occurs around $\Rey\approx 92$. Right: Contours of the cost function $\gamma^2{(k,\omega_n)}$ for divisions of the minima at (d) $\Rey=91$ and (e, f) $\Rey=92$ where the discontinuity occurs. The global optimal actuator positions are marked by a blue dot ($\protect\bluedot$).}
    \label{fig:lqr_optlocsh2n}
\end{figure}

To locate the optimal actuator positions at different Reynolds numbers, we employ the same gradient minimisation method as that used in the OE problem with sensible initial guesses chosen based on the results of figure \ref{fig:lqr_coarse_cont}. The corresponding optimal placements and their control performances are summarised in figure \ref{fig:lqr_optlocsh2n}. We see that the optimal actuator position ($x_{a-opt}$, $y_{a-opt}$) moves downstream and closer to the recirculation zone as the Reynolds number increases. This trend is the opposite of that observed for a spatially developing one-dimensional flow where the optimal actuator placement moves upstream with increasing instability \citep{oehler2018sensor}. This occurs because the absolutely unstable region, as shown in figure \ref{fig:lqe_optlocsh2n}(c), extends only further downstream when the instability increases, whereas it also expands upstream in the one-dimensional Ginzburg-Landau equation. The cost function $\gamma^2{(k,\omega_n)}$ for the FIC controller design and the optimal control performance $\textbf{\textit{J}}_\textrm{FIC}$ from numerical simulations are plotted as a function of the Reynolds number in figure \ref{fig:lqr_optlocsh2n}(c). Both quantities increase logarithmically with increasing Reynolds number and the gap between them represents the contribution from the `background' or `free-stream' modes that are not important for the optimal controller design (see Appendix \ref{sec:app.b}).

Another interesting feature observed is that there exists a discontinuity in the optimal actuator location around $\Rey\approx 92$, as indicated by the dotted line in figure \ref{fig:lqr_optlocsh2n}(a, b). To clearly show why it occurs, we plot contours of the cost function near the optimal positions at $\Rey=91$ and $\Rey=92$ in figure \ref{fig:lqr_optlocsh2n}(d, e). First of all, we identify that one global minimum dominates the near-wake area for $\Rey\leq 90$ but that it splits into two local minima at $\Rey=91$, as shown in \ref{fig:lqr_optlocsh2n}(d). The blue dot ($\protect\bluedot$) marks the global optimal actuator location which still follows the trend of the optimal actuator placement before the discontinuity occurs. Second, these local minima move relatively far from each other at $\Rey=92$ and thus we show them separately using two panels in figure \ref{fig:lqr_optlocsh2n}(e, f). In this case, the global optimum, as marked by the blue dot ($\protect\bluedot$), switches to the one with the lower transverse position which is far downstream and follows this path at higher Reynolds numbers. Note that at a slightly higher Reynolds number, e.g.~$\Rey=93$, the local minimum shown in (e) disappears and only the downstream minimum in panel (f) remains. Third, even though the optimal position changes rapidly near $\Rey=92$, any corresponding discontinuity in the control performance is barely observable in figure \ref{fig:lqr_optlocsh2n}(c). In both cases, the difference in the control performance between these two local minima is small ($<2\%$). A similar discontinuity has also been observed in the spacing and the frequency of vortex shedding around $\Rey\approx 90$ which reveals a small transition of the wake's instability \citep{tritton_1959,lienhard1966synopsis}.

\subsection{Collocated input-output control problem}
\subsubsection{Brute-force sampling}\label{sec:lqg_bfs}
\begin{figure}
    \centerline{
    \hspace{2mm}
    \includegraphics[width=0.975\textwidth]{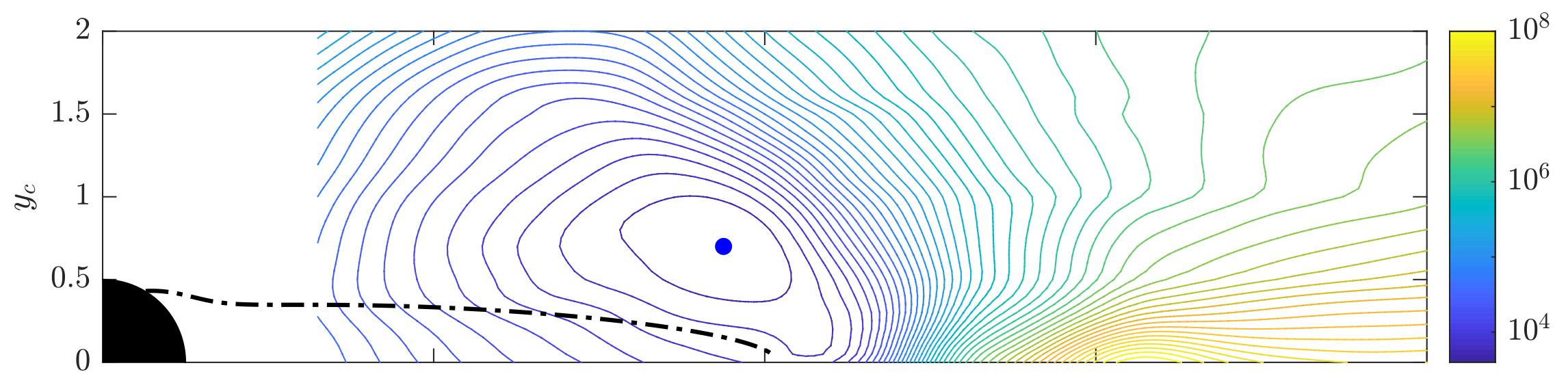}
    \llap{\parbox[b]{5.30in}{(a)\\\rule{0ex}{1.1in}}}
    }
    \centerline{
    \hspace{2mm}
    \includegraphics[width=0.975\textwidth]{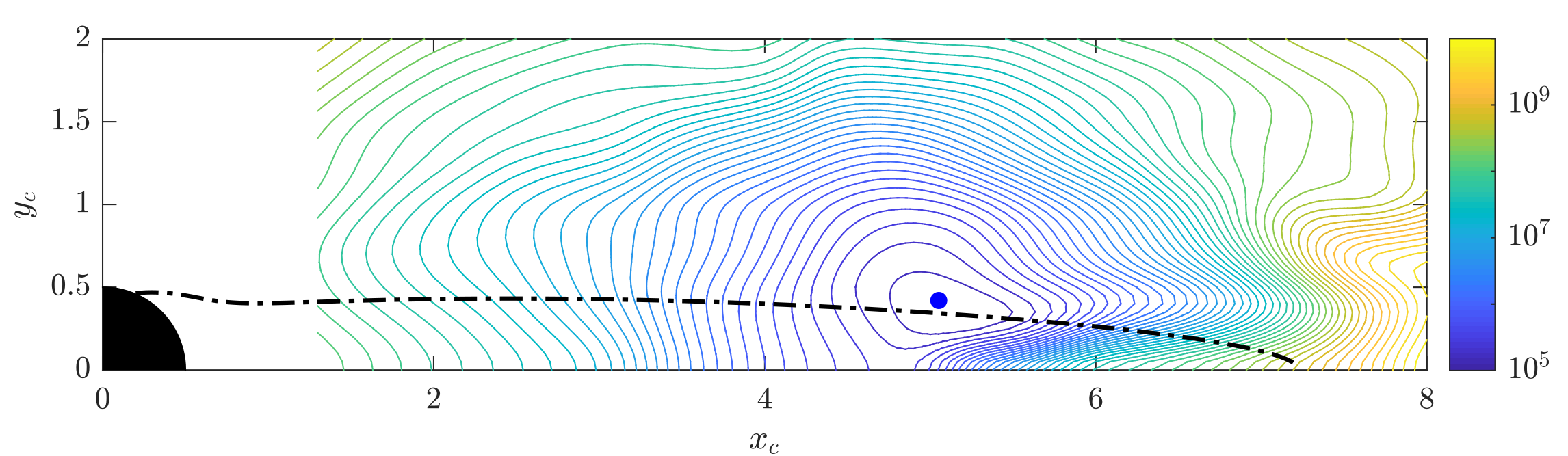}
    \llap{\parbox[b]{5.30in}{(b)\\\rule{0ex}{1.3in}}}
    }
    \caption[Brute-force sampling of the collocated actuator-sensor pair in the CIOC problem]{Contours of the cost function $\gamma^2{(k,\omega_n)}$ in the CIOC problem and the optimal locations of the collocated actuator-sensor pair ($\protect\bluedot$) at (a) $\Rey=60$ and (b) $\Rey=110$. Feedback controllers are designed using the parameters $k=3$ and $\omega_n=9$. The dash-dotted line (\protect\blacklineshort\hspace{0.5mm}\protect\blacksmalldot\hspace{0.5mm}\protect\blacklineshort) indicates the boundary of the reverse-flow region.}
    \label{fig:lqg_coarse_cont}
\end{figure}
We now look at the collocated input-output control (CIOC) problem depicted in figure \ref{fig:controlsetups}(c), where a single collocated actuator-sensor pair is available for measurement and control. The results of the OE and FIC problems have shown success in solving optimal placement problems and help us to understand the challenges of sensor and actuator placement. In this case, a single actuator and a single sensor are placed together such that a localised feedback loop is formed. Brute-force sampling for the CIOC problem is performed at $\Rey = 60$ and $\Rey = 110$ with the parameters $k=3$ and $\omega_n=9$. The corresponding cost function $\gamma^2{(k,\omega_n)}$ is mapped as a function of the location of the actuator-sensor pair ($x_c$, $y_c$) in figure \ref{fig:lqg_coarse_cont}, where the dash-dotted line indicates the reverse-flow region.

As can be seen from figure \ref{fig:lqg_coarse_cont}, the optimal position for the collocated actuator-sensor pair occurs at approximately  (${x_c,\ y_c}$)=($3.75,\ 0.70$) at $\Rey=60$ and at approximately (${x_c,\ y_c}$)=($5.05,\ 0.42$) at $\Rey=110$. Analogous to the OE and FIC problems, only a single local minimum appears in the sampled area at each Reynolds number, which is therefore the global optimum for the optimal placement problem of a collocated actuator-sensor pair. With increasing Reynolds number, this optimum moves downstream and closer to the edge of the reverse-flow region. Another significant observation from figure \ref{fig:lqg_coarse_cont} is that a `cliff' appears near $x_c\approx 5$ and along the centreline at both Reynolds numbers, where the cost function increases rapidly. This is due to the actuator being placed at a position with small receptivity to momentum forcing and a right half-plane (RHP) zero occurs near the unstable pole in the $\textbf{\textit{q}}$-to-$\textbf{\textit{y}}$ transfer function, which imposes a severe limitation on the actuator's ability to control.

\subsubsection{Optimal actuator-sensor placement}
\begin{figure}
    \center{
    \includegraphics[width=0.48\textwidth,valign=b]{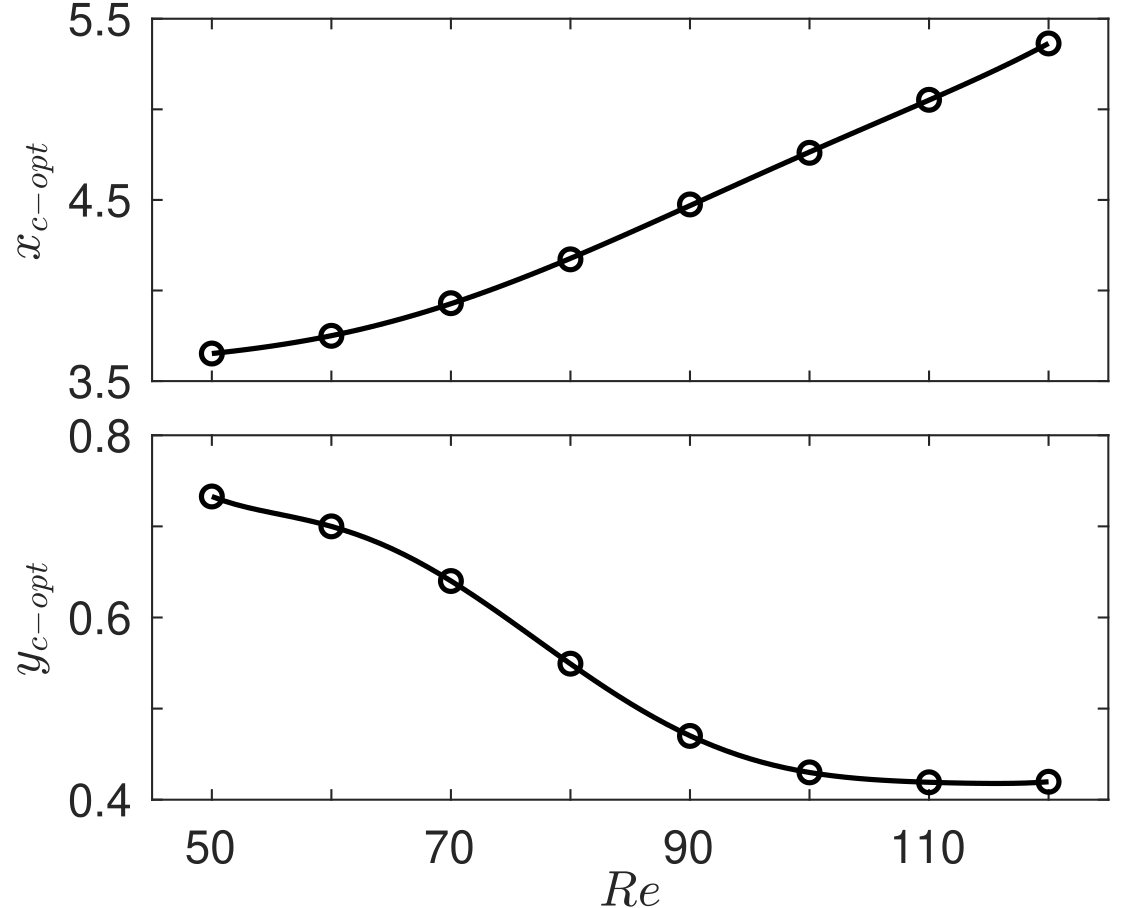}
    \llap{\parbox[b]{2.615in}{(a)\\\rule{0ex}{1.925in}}}
    \includegraphics[width=0.48\textwidth,valign=b]{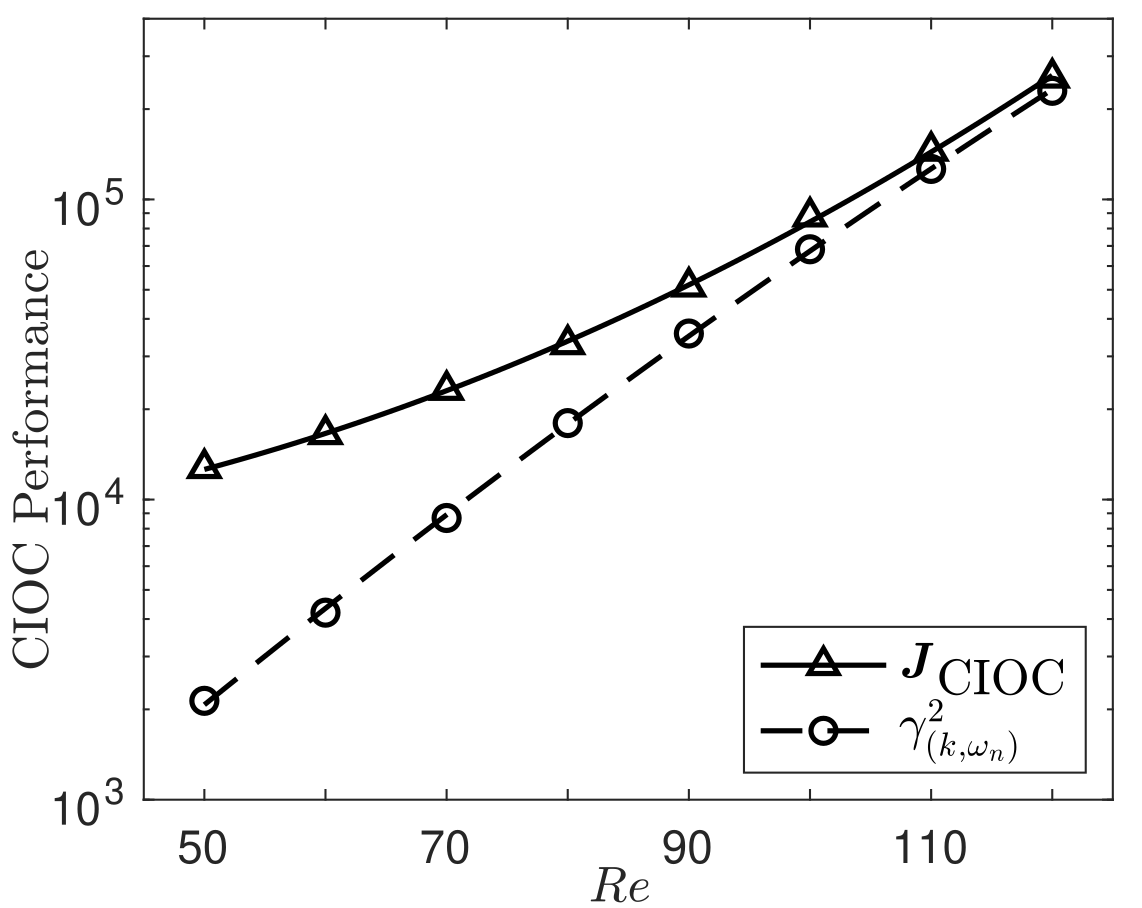}
    \llap{\parbox[b]{5.225in}{(b)\\\rule{0ex}{0.95in}}}
    \llap{\parbox[b]{2.65in}{(c)\\\rule{0ex}{1.925in}}}
    }
    \caption{Coordinates of the optimal sensor-actuator location (a) $x_{c-opt}$ and (b) $y_{c-opt}$ as a function of the Reynolds number. (c) The optimal feedback control performance as a function of the Reynolds number. $\gamma^2{(k,\omega_n)}$ is the cost function for the feedback controller design ($k=3,\ \omega_n=9$) whereas $\textbf{\textit{J}}_\textrm{CIOC}$ is the mean kinetic energy of flow perturbations evaluated from numerical simulations with random disturbances applied everywhere.}
    \label{fig:lqg_optlocsh2n}
\end{figure}

The optimal positions of the collocated actuator-sensor pair are found using the same gradient minimisation method described in \S\ref{sec:lqe} with sensible initial guesses at different Reynolds numbers. The results are displayed in figure \ref{fig:lqg_optlocsh2n} (left panel). The optimal location moves downstream with increasing Reynolds number, which is consistent with the trends observed for both the OE and FIC problems. Meanwhile, the optimal transverse position $y_{c-opt}$, as shown in figure \ref{fig:lqg_optlocsh2n}(b), decreases and eventually converges to a constant position at higher Reynolds numbers. Both the cost function $\gamma^2{(k,\omega_n)}$ and the optimal control performance $\textbf{\textit{J}}_\textrm{CIOC}$ are summarised in figure \ref{fig:lqg_optlocsh2n}(c). In particular, they rise approximately logarithmically with increasing Reynolds number and the gap between them becomes increasingly negligible at higher Reynolds numbers. This is because the contributions from the neglected  `background' or `freestream' modes remain approximately constant regardless of any control, as concluded from figure \ref{fig:lqe_valid_closedsigs}. As a result, they account for a smaller proportion of the total kinetic energy of flow perturbations at higher Reynolds numbers. The comparisons of optimal placements and their performances found for the OE, FIC and CIOC problems will be presented in \S\ref{sec:compares_ofc}.

\subsection{Trade-offs for optimal placement}\label{sec:further}
We now compare the main results of the OE, FIC and CIOC problems, and discuss the implications for effective input-output (feedback) control. Any fundamental trade-offs and key factors that limit effective estimation and control will be explored. In particular, the coupling effect between the sensor and the actuator (i.e.~the time lag) is important due to the convective nature of the flow, and we will highlight the decisive influence of excessive time delay on the sensor and actuator placements. To better demonstrate this, we compare three different IOC setups: i) using an optimally placed collocated actuator-sensor pair (i.e.~CIOC); ii) using optimal sensor and actuator placements found for the OE and FIC problems separately (i.e.~IOC$_{sep}$); iii) using an optimally placed actuator and an optimally placed sensor (i.e.~IOC$_{opt}$). Sensor and actuator placements and their performances for each problem are plotted as a function of the Reynolds number in figure \ref{fig:h2norm_compare_all}(a, b). Details of the problem setups and their optimal performances achieved are then summarised in the table beneath.

\subsubsection{Comparisons of OE, FIC and CIOC problems}\label{sec:compares_ofc}
\begin{figure}
    \begin{minipage}[b]{1.0\textwidth}
        \centering
        \includegraphics[width=0.975\textwidth]{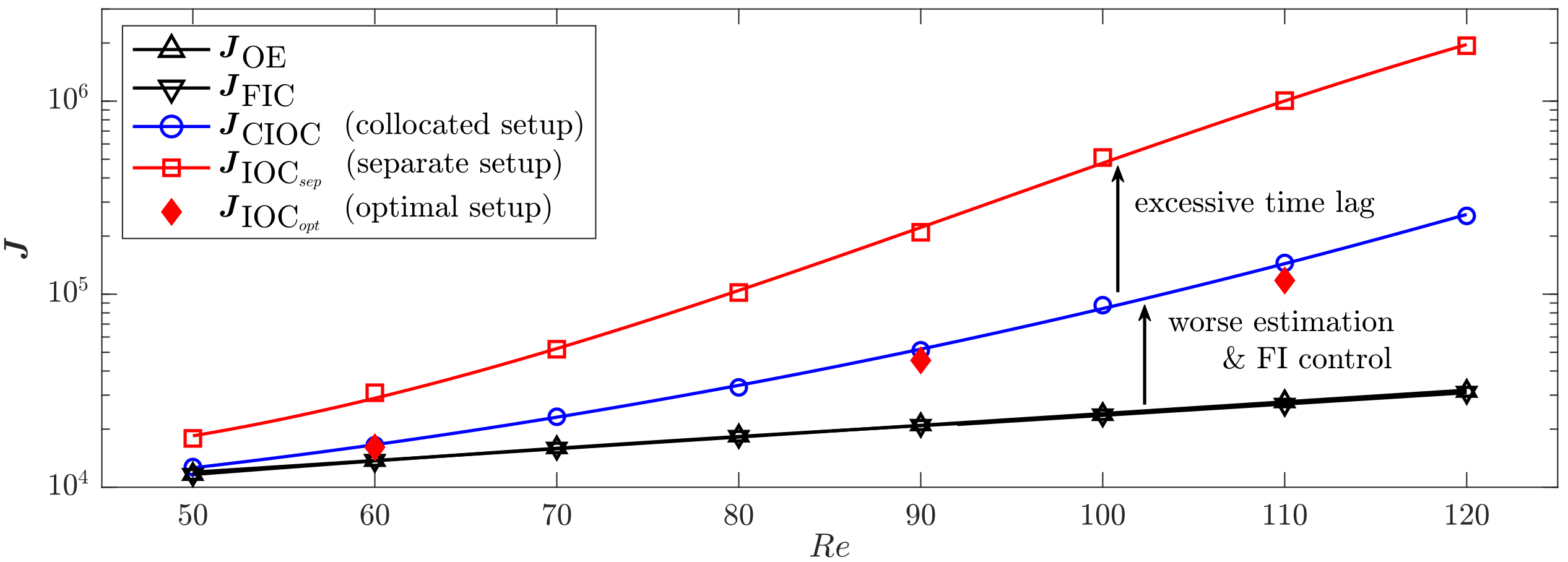}
        \llap{\parbox[b]{5.275in}{(a)\\\rule{0ex}{1.75in}}}
    \end{minipage}
    \begin{minipage}[b]{1.0\textwidth}
        \centering
        \hspace{0.5mm}
        \includegraphics[width=0.4825\textwidth]{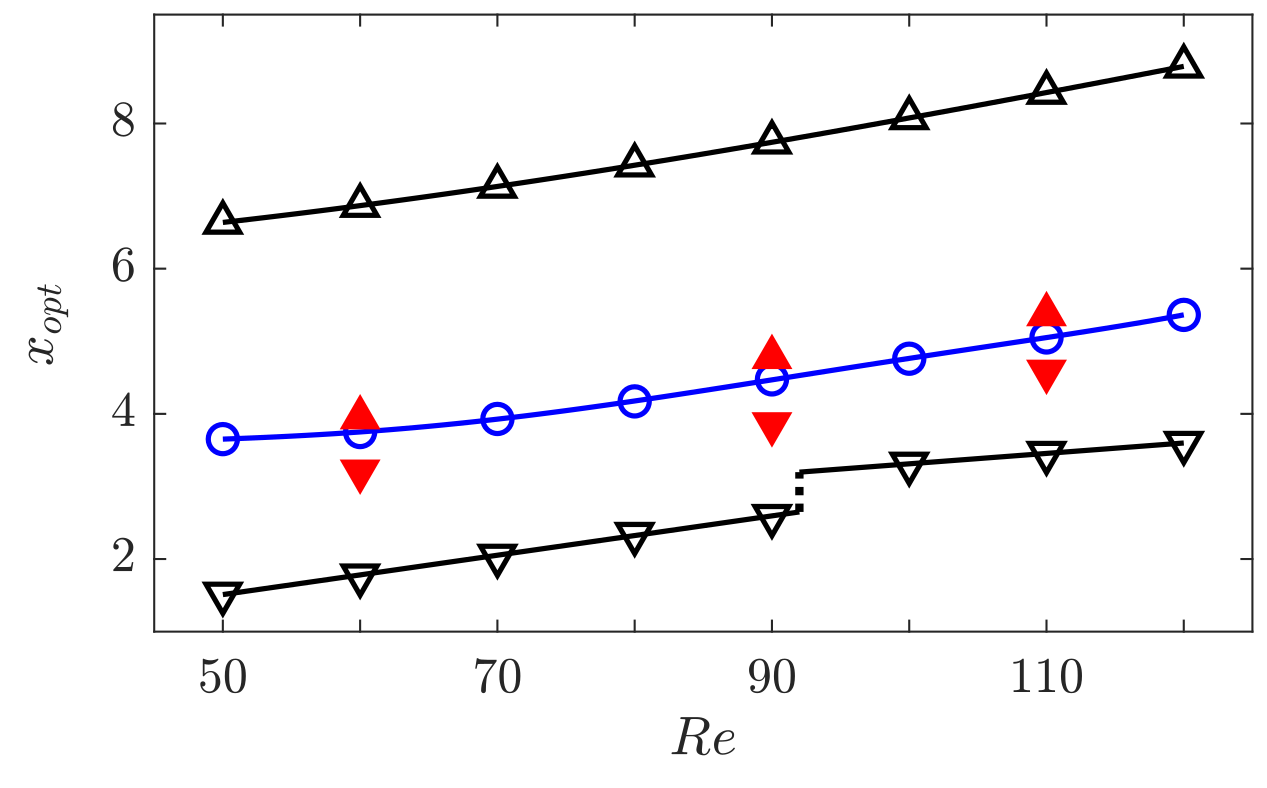}
        \includegraphics[width=0.4825\textwidth]{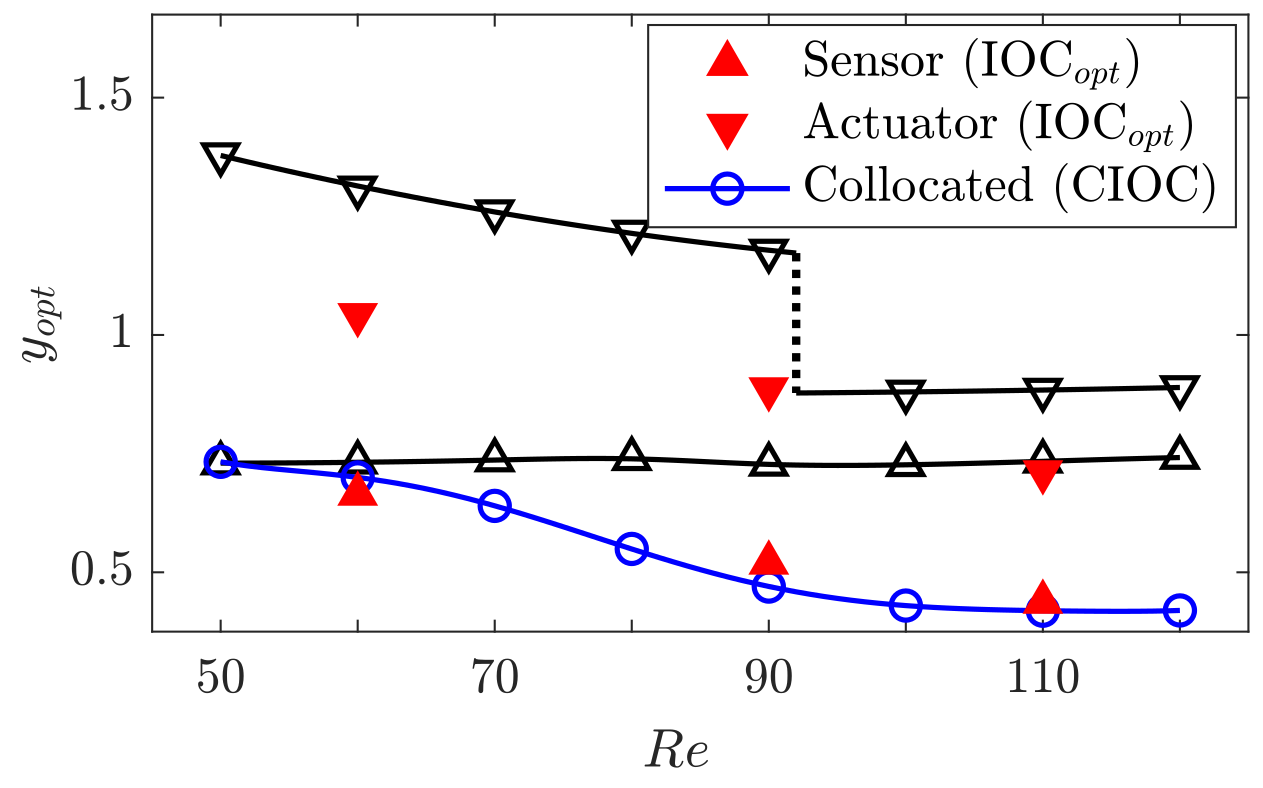}
        \llap{\parbox[b]{5.275in}{(b)\\\rule{0ex}{1.5in}}}
        \vspace{-4mm}
    \end{minipage}
    \begin{minipage}[b]{1.0\textwidth}
    \llap{\parbox[b]{0in}{(c)\\\rule{0ex}{0.25in}}}
    \vspace{-5mm}
        \begin{center}
        \def~{\hphantom{0}}
        \begin{tabular}{lccccccc}
        \toprule
        \multirow{2}{*}{\ \ Problem} &
        \multicolumn{3}{c}{Placement} &&
        \multicolumn{3}{c}{Performance \textbf{\textit{J}} ($\times 10^4$)} \\ \cmidrule{2-4} \cmidrule{6-8}
        &
        \begin{tabular}[c]{@{}c@{}}\ \ \ estimation\ \ \ \\ (sensor)\end{tabular} &
        \begin{tabular}[c]{@{}c@{}}\ \ \ FI control\ \ \ \\ (actuator)\end{tabular} &
        \begin{tabular}[c]{@{}c@{}}\ \ \ time lag\ \ \ \\ (coupling)\end{tabular} &&
        $\Rey$=60 &
        $\Rey$=90 &
        $\Rey$=110 \\ \midrule
        \begin{tabular}[l]{@{}c@{}}OE\hspace{7mm}\ (\dtriangleline{black})\end{tabular} &
        optimal &
        full-state &
        minimal &&
        $1.38$ &
        $2.11$ &
        $2.78$ \\ 
        \begin{tabular}[l]{@{}c@{}}FIC\hspace{7mm}(\triangleline{black})\end{tabular} &
        full-state &
        optimal &
        minimal &&
        $1.37$ &
        $2.08$ &
        $2.68$ \\ 
        \begin{tabular}[l]{@{}c@{}}CIOC\hspace{4.5mm}(\cirline{blue})\end{tabular} &
        --- &
        --- &
        minimal &&
        $1.65$ &
        $5.12$ &
        $14.51$ \\ 
        \begin{tabular}[l]{@{}c@{}}IOC$_{sep}$\hspace{2.1mm}(\squareline{red})\end{tabular} &
        optimal &
        optimal &
        --- &&
        $3.09$ &
        $20.92$ &
        $100.59$ \\ 
        \begin{tabular}[l]{@{}c@{}}IOC$_{opt}$\hspace{2.1mm}(\ \ \diamondfill{red}\ \ )\end{tabular} &
          \multicolumn{3}{c}{best trade-off between all} &&
          $1.61$ &
          $4.54$ &
          $11.77$ \\ \bottomrule
        \end{tabular}
        \end{center}
    \end{minipage}
  \caption[Comparisons of the optimal performances and the optimal placements between different problem setups.]{Comparisons of the optimal performances and the optimal placements between different problem setups. (a) The optimal performance $\textbf{\textit{J}}$ as a function of the Reynolds number in the OE (\dtriangleline{black}), FIC  (\triangleline{black}), CIOC (collocated setup, \cirline{blue}), IOC$_{sep}$ (separate setup, \squareline{red}) and IOC$_{opt}$ (optimal setup, \diamondfill{red}) problems. (b) Coordinates of optimal placements as a function of the Reynolds number. (c) A summary of setups for\protect\\all problems and their optimal performances at $\Rey$=60, 90, 110.}
\label{fig:h2norm_compare_all}
\end{figure}

We first consider the OE problem (represented by \triangleline{black}) and the FIC problem (represented by \dtriangleline{black}), in which the sensor and actuator are each placed to achieve the best estimation and FI control performance possible, respectively. From figure \ref{fig:h2norm_compare_all}(a), we first notice that the optimal OE performance $\textbf{\textit{J}}_\textrm{OE}$ (\triangleline{black}) and the optimal FIC performance $\textbf{\textit{J}}_\textrm{FIC}$ (\dtriangleline{black}) are almost identical for each Reynolds number considered. As aforementioned, the OE problem can be recast as a unity feedback control problem for the estimation error with perfect actuation, and the optimal sensor placement strikes a balance between measuring the flow upstream and measuring the flow downstream to provide the best observation of the entire flow. Similarly, the FIC problem provides perfect measurements of the entire flow field, and the optimal actuator placement achieves the best FIC performance by striking a balance between attenuating flow perturbations upstream and attenuating flow perturbations downstream. The differences between the OE and FIC problems are summarised in figure \ref{fig:h2norm_compare_all}(c), and the similarities between their optimal performances implies that neither the sensor placement nor the actuator placement is the key factor that limits control performance. It is also interesting to note that there exists a large spatial separation between the optimal sensor placements for the OE problem (\triangleline{black}) and the optimal actuator placements for the FIC problem (\dtriangleline{black}), as shown in figure \ref{fig:h2norm_compare_all}(b). This is mainly caused by the convective non-normality of the cylinder flow, which leads to upstream-leaning forcing modes and downstream-leaning response modes.

We further consider the CIOC problem (represented by \cirline{blue}) in which the minimal possible time lag between actuation and sensing is achieved. In this scenario, the optimal placement of the collocated actuator-sensor pair ensures the best trade-off between maintaining good estimation performance and maintaining good FI control performance. We can see that $\textbf{\textit{J}}_\textrm{CIOC}$ (\cirline{blue}) is significantly larger than either $\textbf{\textit{J}}_\textrm{OE}$ or $\textbf{\textit{J}}_\textrm{FIC}$, particularly at higher Reynolds numbers, as shown in figure \ref{fig:h2norm_compare_all}(a). This performance deterioration is mainly caused by two factors: i) the perfect spatially distributed actuation (in the OE problem) and the perfect spatially distributed sensing (in the FIC problem) degenerate to a single-point actuator and a single-point sensor in the CIOC problem; ii) neither the sensor nor the actuator is optimally placed, both of which contributing to poorer estimation and control of the entire flow than for the OE and FIC problems separately. As shown in figure \ref{fig:h2norm_compare_all}(b), the optimal placement of the collocated actuator-sensor pair (\cirline{blue}) is located between the optimal placements found for the OE and FIC problems, which is expected behaviour. 

\subsubsection{IOC using optimal placements of OE and FIC}\label{sec:compares_ci}
The input–output controller that results from the independently designed optimal estimator and the independently designed optimal full-state information controller is still optimal, as stated by the separation principle of estimation and control. However, the optimal placements found for the OE and FIC problems are not necessarily optimal for the input-output control (IOC) problem. That is, the optimal placement problem does not satisfy the separation principle. To better demonstrate this, we now consider the IOC problem that uses a setup where a single sensor is placed at the optimal location found for the OE problem to provide the best estimation of the entire flow and a single actuator is placed at the optimal location found for the FIC problem to provide the best FI control of the entire flow. This setup utilises optimal sensor and actuator locations that were found independently for the OE and FIC problems, and is thus denoted by IOC$_{sep}$ (\squareline{red}) in figure \ref{fig:h2norm_compare_all}. Although the best estimation performance and the best FI control performance are each ensured, the corresponding feedback control (IOC) performance exhibits a more severe deterioration when compared to the CIOC problem, as shown in figure \ref{fig:h2norm_compare_all}(a). In particular, the cost function $\textbf{\textit{J}}_{\textrm{IOC}_{sep}}$ is 87$\%$ higher than $\textbf{\textit{J}}_\textrm{CIOC}$ at $\Rey=60$ and it is 593$\%$ higher at $\Rey=110$, as listed by the table in figure \ref{fig:h2norm_compare_all}(c). 

As discussed in \S\ref{sec:lqe_domain_size}, one possible reason for the performance deterioration is that the sensor in the OE problem and the sensor in the feedback control problem are required to measure different information to achieve their respective best performance. Although the optimal sensor placement found for the OE problem provides the best estimation of the entire flow field (e.g.~both instability activities and transportation of flow perturbations), it fails to optimally estimate the information that would be the most beneficial for feedback control. Another significant reason is the excessive time lag between the sensor and the actuator, which has a much stronger influence for the IOC problem. This is caused by the convective nature of the cylinder flow. The sensor and actuator placements that are most effective for feedback control should therefore give priority to reducing the time lag between actuation and sensing. 

\subsubsection{Optimal IOC and the effect of time delay}\label{sec:compares_ci_opt}
The effect of time lag can be more clearly shown by considering the optimal setup for the IOC problem, in which an optimally placed sensor and an optimally placed actuator are used to achieve the best IOC performance. It is computationally expensive to find such placements for a two-dimensional flow since the optimisation problem is four-dimensional (locating the optimal sensor and optimal actuator placements simultaneously). Therefore, we only present results for $\Rey=60,\ \Rey=90\ \text{and}\ \Rey=110$. These results are represented by IOC$_{opt}$ (\diamondfill{red}) in figure \ref{fig:h2norm_compare_all}. As can be seen from figure \ref{fig:h2norm_compare_all}(a), the control performance $\textbf{\textit{J}}_{\textrm{IOC}_{opt}}$ is only slightly smaller than $\textbf{\textit{J}}_\textrm{CIOC}$ (\cirline{blue}). It is approximately 2.5$\%$ lower at $\Rey=60$ and 19$\%$ lower at $\Rey=110$, as listed in figure \ref{fig:h2norm_compare_all}(c). 

We also notice that the optimal sensor and actuator placements in the IOC$_{opt}$ problem (\mytriangle{red}/\mydtriangle{red}) are close to those found for the CIOC problem (\cirline{blue}), as shown in figure \ref{fig:h2norm_compare_all}(b). The optimal placement of a collocated actuator-sensor pair therefore seems to provide a good approximation of the optimal feedback control setup for the cylinder flow. In particular, the optimal sensor locations (\mytriangle{red}) for feedback control show good agreement with the optimally placed collocated actuator-sensor pair (\cirline{blue}) in the CIOC problem at both Reynolds numbers. As for the optimal actuator positions (\mydtriangle{red}), their streamwise coordinates (i.e.~$x_{opt}$) are slightly upstream of those found for the CIOC problem (\cirline{blue}) whereas their transverse coordinates lie between those found for the FIC problem and those found for the CIOC problem. 

This observation is not consistent with the intuition put forward by many previous studies that maintaining accurate flow estimation and maintaining effective FI control are essential for efficient feedback control \citep{cohen2006heuristic,seidel2009feedback}. Specifically, the best sensor location for feedback control should be close to that found for the OE problem and the best actuator location for feedback control should be close to that found for the FIC problem. This is true in some previous studies for simpler spatially-developing flow models, e.g.~for the one-dimensional Ginzburg-Landau equation \citep{oehler2018sensor}. However, the current study reveals that minimising the time-lag effect is of higher priority for feedback control of the two-dimensional cylinder flow. One possible explanation for the consistency between the optimal placements found for the CIOC problem and those found for the IOC problem is the convective nature of the cylinder flow which leads to a strong time-lag effect between the separately placed sensor and actuator. Indeed, by introducing an artificial time delay into the one-dimensional Ginzburg-Landau equation, the optimal sensor and actuator locations in the feedback control problem move closer to each other and lead to similar results as those shown in the current study \citep{oehler2018sensor}. This supports the observation that the time-lag is indeed the major factor that determines the optimal placement of control devices in the two-dimensional cylinder flow.

\section{Conclusions}\label{sec:conclusions}
We have considered optimal estimation and control of linear perturbations in the flow past a two-dimensional circular cylinder over a range of Reynolds numbers. In particular, we focused on the optimal placement of a single sensor and a single actuator to better understand the limitations of effective feedback control. Although the corresponding optimal placement problems are non-convex, brute-force sampling results for each problem revealed a unique local optimal position in the wake area which represents the global optimum. A simple gradient minimisation method was then sufficient to locate the optimal positions at each Reynolds number. It was shown that the optimal sensor and actuator locations move downstream as Reynolds number increases, and their trajectories presented conflicting trade-offs. In the OE problem, the sensor should be placed where it achieves the best compromise between measuring the flow upstream and measuring the flow downstream to provide optimal observations of the entire flow. In the FIC problem, the optimal actuator placement should strike a balance between controlling the near-wake area (which displays receptivity to disturbances) and controlling the far-wake area where the remaining disturbances can potentially be amplified. 

For the input-output (feedback) control problem in which the sensor and actuator placements are coupled, the effect of excessive time lag results in optimal sensor and actuator locations that are close to each other in the streamwise direction. In particular, the optimal placement of a collocated actuator-sensor pair was shown to be a good approximation for the actual optimal placements for feedback control. For fluid flows that are dominated by convection, e.g.~the cylinder flow, reducing the time lag between actuation and sensing appears to be crucial for achieving good feedback control performance. 

\appendix
\section{The systems for estimation and control}\label{sec:app.a}
Table \ref{tab:app.sysmats_summary} lists the system states, inputs, outputs, and Riccati equations for the OE, FIC and IOC problems. Note that the state-space model of the IOC problem is assembled from two subsystems of the OE and FIC problems. The corresponding state matrices are $\Tilde{\textbf{E}}=\textbf{diag}[\textbf{E},\textbf{E}]$ and $\Tilde{\textbf{A}}=\textbf{diag}[\textbf{A},\textbf{A}]$, where $\textbf{diag}[\cdot]$ indicates a block-diagonal matrix built from the provided matrices. Therefore, all three problems can be cast into the same general form \eqref{equ:model_for_all} \citep{kim2007linear,skogestad2007multivariable,chen2011h}:
\begin{equation}\label{equ:model_for_all}
    \begin{bmatrix}
        \Tilde{\textbf{E}}\dot{\textbf{\textit{x}}}\\
        \textbf{\textit{y}}\\
        \textbf{\textit{z}}
    \end{bmatrix}=
    \begin{bmatrix}
        \Tilde{\textbf{A}} & \textbf{B}_1 & \textbf{B}_2 & \textbf{0}\\
        \textbf{C}_1 & \textbf{0} & \textbf{0} & \textbf{V}^{1/2}\\
        \textbf{C}_2 & \textbf{R}^{1/2} & \textbf{0} & \textbf{0}
    \end{bmatrix}
    \begin{bmatrix}
        \textbf{\textit{x}}\\
        \textbf{\textit{q}}\\
        \textbf{\textit{d}}\\
        \textbf{\textit{n}}
    \end{bmatrix}
\end{equation}
In particular, the optimal estimator gain $\textbf{K}_f$ and the FI control gain $\textbf{K}_r$ are formed from the solutions of the algebraic Riccati equations associated with the OE and FIC problems. The covariance matrices for the OE and FIC problems are $\textbf{W}_1=\textbf{B}_d\textbf{B}_d^T$ and $\textbf{W}_2=\textbf{C}_z^T\textbf{C}_z$, respectively. In this case, $\textbf{B}_d$ and $\textbf{C}_z$ are the low-rank input and output matrices (see \S\ref{sec:design_method}) to overcome the difficulty of solving Riccati equations with full-state inputs and full-state outputs. 

\begin{table}
  \begin{center}
\def~{\hphantom{0}}
\begin{tabular}{c|ccc|ccc|cc|cc}
\hline
\multirow{2}{*}{P} &
  \multicolumn{3}{c|}{State} &
  \multicolumn{3}{c|}{Input} &
  \multicolumn{2}{c|}{Output $\textbf{\textit{y}}$} &
  \multicolumn{2}{c}{Performance $\textbf{\textit{z}}$} \\ 
& \textbf{\textit{x}} & $\Tilde{\textbf{E}}$ & $\Tilde{\textbf{A}}$ & \textbf{\textit{q}} & $\textbf{B}_1$  & $\textbf{B}_2$ & $\textbf{C}_1$& \multicolumn{1}{c|}{$\textbf{V}^{1/2}$}& $\textbf{C}_2$& $\textbf{R}^{1/2}$\\ \hline
\multirow{4}{*}{OE} & 
\multirow{4}{*}{\textbf{\textit{e}}} & \multirow{4}{*}{\textbf{E}} & \multirow{4}{*}{\textbf{A}} & 
\multirow{2}{*}{-\textbf{\textit{y}}} & 
\multirow{2}{*}{$\textbf{K}_f$} & 
\multirow{2}{*}{$\textbf{B}_d$} & 
\multirow{2}{*}{$\textbf{C}_y$} & 
\multicolumn{1}{c|}{\multirow{2}{*}{$\alpha\textbf{I}$}} & 
\multirow{2}{*}{$\textbf{C}_z$} & 
\multirow{2}{*}{$\textbf{0}$} \\
& & & & & & & & \multicolumn{1}{c|}{} & & \\
& & & & \multicolumn{5}{c}{\multirow{2}{*}{$\textbf{E}\textbf{X}\textbf{A}^T+\textbf{A}\textbf{X}\textbf{E}-\textbf{E}\textbf{X}\textbf{C}_y^T\textbf{V}^{-1}\textbf{C}_y\textbf{X}\textbf{E}+\textbf{W}_1=0$}} & \multicolumn{2}{c}{\multirow{2}{*}{$\textbf{K}_f=\textbf{E}\textbf{X}\textbf{C}_y^T\textbf{V}^{-1}$}} \\ 
& & & & \multicolumn{5}{c}{} & \multicolumn{2}{c}{} \\ \cline{5-11}
\multirow{4}{*}{FIC} & \multirow{4}{*}{\textbf{\textit{w}}} & \multirow{4}{*}{\textbf{E}} & \multirow{4}{*}{\textbf{A}} & \multirow{2}{*}{-\textbf{\textit{y}}} & \multirow{2}{*}{$\textbf{B}_q$}& \multirow{2}{*}{$\textbf{B}_d$} & \multirow{2}{*}{$\textbf{K}_r$} & \multicolumn{1}{c|}{\multirow{2}{*}{$\textbf{0}$}} & \multirow{2}{*}{$\textbf{C}_z$} & \multirow{2}{*}{$\beta\textbf{I}$} \\   
& & & & & & & & \multicolumn{1}{c|}{} & & \\ 
& & & & \multicolumn{5}{c}{\multirow{2}{*}{$\textbf{E}^T\textbf{Y}\textbf{A}+\textbf{A}^T\textbf{Y}\textbf{E}-\textbf{E}^T\textbf{Y}\textbf{B}_q\textbf{R}^{-1}\textbf{B}_q^T\textbf{Y}\textbf{E}+\textbf{W}_2=0$}} & \multicolumn{2}{c}{\multirow{2}{*}{$\textbf{K}_r=\textbf{R}^{-1}\textbf{B}_q^T\textbf{Y}\textbf{E}$}} \\ 
& & & & \multicolumn{5}{c}{} & \multicolumn{2}{c}{} \\ \cline{5-11}
\multirow{3}{*}{IOC} &
  \multirow{3}{*}{\begin{tabular}[c]{@{}c@{}}
    $\begin{bmatrix}
    \textbf{\textit{w}} \\ \textbf{\textit{e}}
    \end{bmatrix}$
  \end{tabular}} & 
  \multirow{3}{*}{\begin{tabular}[c]{@{}c@{}}
    $\begin{bmatrix}
    \textbf{E} & \textbf{0}\\
    \textbf{0} & \textbf{E}
    \end{bmatrix}$
  \end{tabular}} & 
  \multirow{3}{*}{\begin{tabular}[c]{@{}c@{}}
    $\begin{bmatrix}
    \textbf{A} & \textbf{0}\\
    \textbf{0} & \textbf{A}
    \end{bmatrix}$
  \end{tabular}} & 
  \multirow{3}{*}{-\textbf{\textit{y}}} &
  \multirow{3}{*}{\begin{tabular}[c]{@{}c@{}}
    $\begin{bmatrix}
    \textbf{B}_q&\textbf{0}\\ \textbf{0}&\textbf{K}_f
    \end{bmatrix}$
  \end{tabular}} &
  \multirow{3}{*}{\begin{tabular}[c]{@{}c@{}}
    $\begin{bmatrix}
    \textbf{B}_d \\ \textbf{B}_d
    \end{bmatrix}$
  \end{tabular}} &
  \multirow{3}{*}{\begin{tabular}[c]{@{}c@{}}
    $\begin{bmatrix}
    \textbf{K}_r&-\textbf{K}_r\\ \textbf{0}&\textbf{C}_y
    \end{bmatrix}$
  \end{tabular}} &
  \multirow{3}{*}{\begin{tabular}[c]{@{}c@{}}
    $\begin{bmatrix}
    \textbf{0}\\ \alpha\textbf{I}
    \end{bmatrix}$
  \end{tabular}} &
  \multirow{3}{*}{\begin{tabular}[c]{@{}c@{}}
    $\begin{bmatrix}
    \textbf{C}_z&\textbf{0}
    \end{bmatrix}$
  \end{tabular}} &
  \multirow{3}{*}{\begin{tabular}[c]{@{}c@{}}
    $\begin{bmatrix}
    \textbf{0}&\beta\textbf{I}
    \end{bmatrix}$
  \end{tabular}} \\
 & & & & & & & & & & \\
 & & & & & & & & & & \\ \hline
\end{tabular}
\caption[A summary of system states, matrices and  Riccati equations for the OE, FIC and IOC problems.]{A summary of system states, matrices and  Riccati equations for the OE, FIC and IOC problems.}
\label{tab:app.sysmats_summary}
\end{center}
\end{table}

\section{Convergence analysis}\label{sec:app.b}
This section presents the convergence analysis for the optimal performance and the optimal placement concerning the frequency range $\omega_n$ and the number of resolvent modes $k$. The main results are summarised in table \ref{tab:lqe_valid_optlocs}, where the OE problem is solved at $\Rey=90$ for different choices of $\omega_n$ and $k$. Note that the cost function $\gamma^2{(k,\omega_n)}$ is computed from the integration of the first $k$ energy gains ($\sigma_i^2$) of the closed-loop system across the frequency range $\omega\in[-\omega_n,\ \omega_n]$ (i.e.~equation \eqref{equ:h2norm_resovent}) whereas the estimation performance $\textbf{\textit{J}}_\textrm{OE}$ is evaluated directly from numerical simulations when disturbances are applied everywhere in the domain (i.e.~equation \eqref{equ:cost_j}). Any difference between them is accounted for by the less energetic modes that are neglected during the design of the optimal estimator.

We immediately see that for larger values of $k$ and $\omega_n$, the optimal sensor placement ($x_{s-opt},\ y_{s-opt}$) converges to a constant value. The physical meaning of the reduced cost function $\gamma^2{(k,\omega_n)}$ is the mean kinetic energy of the estimation error while only exciting the most energetic physical mechanisms, whereas $\textbf{\textit{J}}_\textrm{OE}$ represents the mean kinetic energy of the total estimation error. As expected, a larger value of $m$, which corresponds to applying more disturbances, gives a larger cost function $\gamma^2{(k,\omega_n)}$. But the total estimation error $\textbf{\textit{J}}_\textrm{OE}$ eventually converges to a constant value since the performance of the estimator has converged to the global optimum (final relative change of $\textbf{\textit{J}}_\textrm{OE}$ is around $10^{-4}$). 
\begin{table}
  \begin{center}
\def~{\hphantom{0}}
  \begin{tabular}{p{1cm}p{1cm}p{1cm}p{0.9cm}p{0.1cm}p{1cm}p{1cm}p{1.7cm}p{1.5cm}}
    \toprule
      \multirow{2}{*}{$\Rey$}& \multicolumn{3}{c}{Input Parameters} & & \multicolumn{4}{c}{Problem Output} \\[2pt] \cmidrule{2-4} \cmidrule{6-9}
      & $k$& $\omega_n$ & $m$ & & $x_{s-opt}$ & $y_{s-opt}$ & $\gamma^2{(k,\omega_n)}$ &$\textbf{\textit{J}}_\textrm{OE}$ \\[2pt]
    \midrule
      90& 1 & 3 & 29 & & 8.13 & 0.78 &1744.25 &  21366.55 \\
      90& 3 & 9 & 176 & & 7.70 & 0.73 & 3605.16 & 21097.16 \\
      90& 9 & 18 & 768 & &7.70 & 0.73 & 5316.43 & 21094.01 \\ 
    \bottomrule
  \end{tabular}
  \caption[Convergence of the estimation performance in the OE problem]{The rank of the reduced disturbance $m$, optimal sensor location of the OE problem ($x_{s-opt}$, $y_{s-opt}$), the cost function $\gamma^2{(k,\omega_n)}$ for the optimal estimator design and the mean energy of the total estimation error $\textbf{\textit{J}}_\textrm{OE}$ (evaluated from DNS with random disturbances applied everywhere) are listed with different parameters ($k$, $\omega_n$) at $\Rey=90$.}
  \label{tab:lqe_valid_optlocs}
  \end{center}
\end{table}
\begin{figure}
    \vspace{2mm}
    \centerline{
    \hspace{0mm}
    \includegraphics[width=0.48\textwidth]{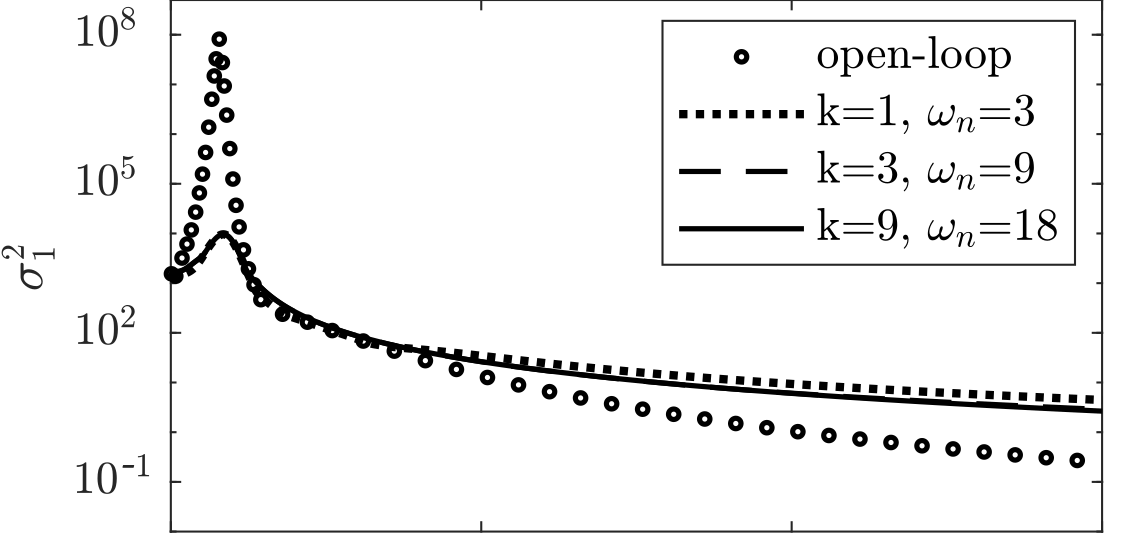}
    \llap{\parbox[b]{2.65in}{(a)\\\rule{0ex}{1.175in}}}
    \hspace{0mm}
    \includegraphics[width=0.48\textwidth]{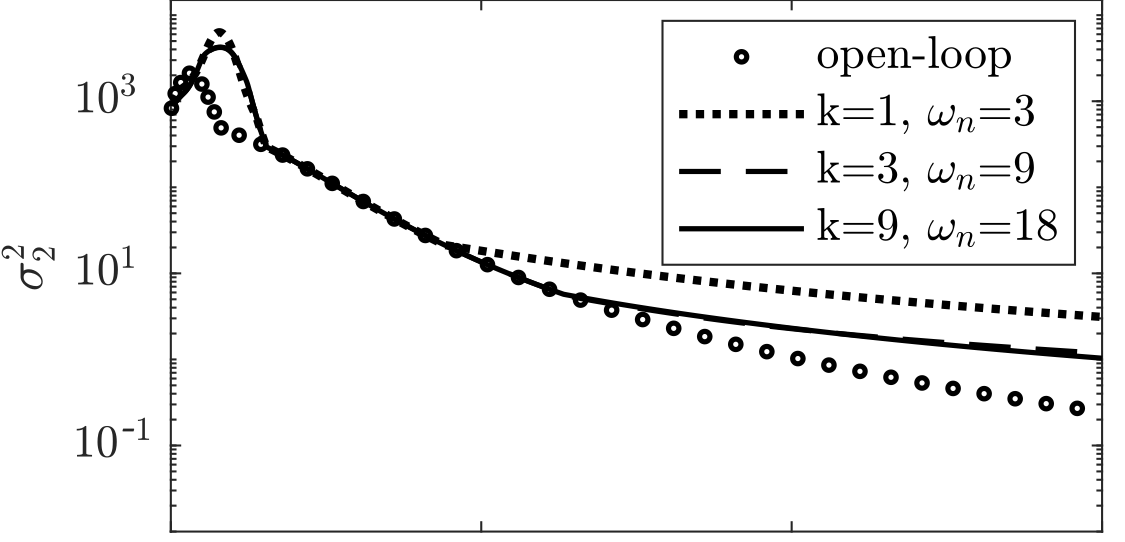}
    \llap{\parbox[b]{2.65in}{(b)\\\rule{0ex}{1.175in}}}
    }
    \vspace{0mm}
    \centerline{
    \hspace{0mm}
    \includegraphics[width=0.48\textwidth]{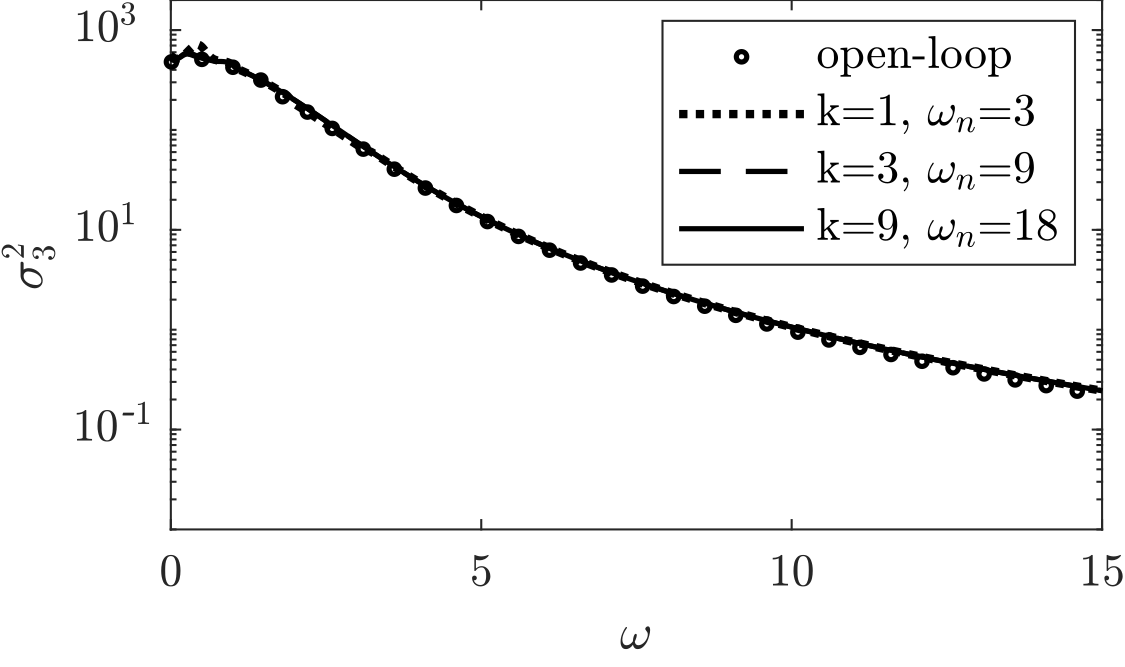}
    \llap{\parbox[b]{2.65in}{(c)\\\rule{0ex}{1.45in}}}
    \hspace{0mm}
    \includegraphics[width=0.48\textwidth]{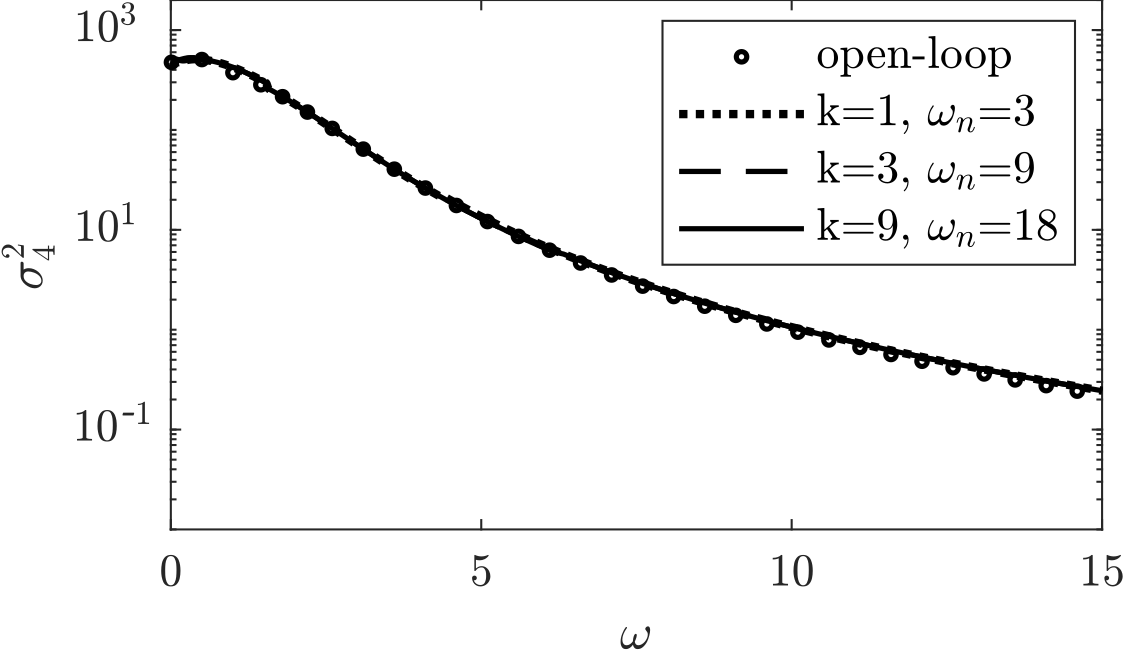}
    \llap{\parbox[b]{2.65in}{(d)\\\rule{0ex}{1.45in}}}
    }
    \caption[Comparisons of the first four resolvent spectra between the open-loop system and the closed-loop error systems]{Comparisons of (a) the first singular value $\sigma_1$, (b) the second singular value $\sigma_2$, (c) the third singular value $\sigma_3$ and (d) the fourth singular value $\sigma_4$ from resolvent analysis of the open-loop system $\textbf{P}(s)$ ($\circ$) and the closed-loop systems $\textbf{G}(s)$ (lines) at $\Rey=90$. Estimators are designed for a sensor placed at $(7.70,\ 0.73)$ for different $k$ and $\omega_n$.}
    \label{fig:lqe_valid_closedsigs}
\end{figure}

Figure \ref{fig:lqe_valid_closedsigs} compares the first four resolvent spectra computed from the open-loop system $\textbf{P}(s)$ (without estimator) to those from the closed-loop error system $\textbf{G}(s)$. The optimal estimators are designed with a sensor placed at the converged optimal location that is listed in table \ref{tab:lqe_valid_optlocs}. In figure \ref{fig:lqe_valid_closedsigs}, dashed lines ($k=3,\ \omega_n=9$) and solid lines ($k=9,\ \omega_n=18$) are perfectly matched so that the resolvent spectra of the closed-loop error system are already converged. The estimator significantly modifies the first two singular values but barely modifies either the third or the fourth singular values. Therefore, it is unnecessary to optimise all energy gains over all frequencies and a reasonable choice of $k$ and $\omega_n$ leads to convergence to the global optimum. The gap between the cost function $\gamma_{(k,\omega_n)}^2$ and $\textit{\textbf{J}}_\textrm{OE}$ is accounted for by the effect of `background' or `freestream' modes which are not important for optimal estimator design or optimal sensor placement. The parameter combination $k=3$ and $\omega_n=9$ is therefore sufficient for the current study. Analogous to the OE problem, these same parameter values ($k=3$ and $\omega_n=9$) are also found to be sufficient for the FIC and IOC problems. 

\section{The root mean square of the norm}\label{sec:app.c}
The root-mean-square value $\epsilon(x,y)$ is defined as:
\begin{equation}\label{equ:app.pngilon}
    \gamma^2(k,\omega)=\dfrac{1}{2\pi}\int_{-\omega_n}^{\omega_n}\sum_{i=1}^{k}\sigma^2_i(j\omega)\ d\omega=\int_{\Omega}\epsilon^2(x,y)\ d\Omega\ ,
\end{equation}
where $\gamma^2(k,\omega)$ is the cost function of the associated design problem. In the OE problem, we can solve for $\epsilon_{\textrm{OE}}^2$ by using resolvent response modes:
\begin{equation}
    \epsilon_{\textrm{OE}}^2=\sum_{u,v}\Big\{\dfrac{1}{2\pi}\int_{-\omega_n}^{\omega_n}\sum_{i=1}^{k}\left(\sigma_i\hat{\textbf{u}}_i\right)^{\odot 2}\ d\omega\Big\}\ ,
\end{equation}
where the singular values $\sigma_i$ and resolvent response modes $\hat{\textbf{u}}_i$ are computed from resolvent analysis of the closed-loop system $\textbf{G}(s)$. Here, $()^{\odot}$ is the Hadamard power and $\sum_{u,v}$ denotes summation over the streamwise and transverse components. In the FIC problem, a similar root-mean-square value $\epsilon_{\textrm{FIC}}$ can be defined from the resolvent forcing modes. 

\bibliographystyle{jfm}
\bibliography{jfm}

\end{document}